\documentclass[journal]{IEEEtran}

\usepackage[dvipsnames]{xcolor}
\usepackage{graphicx}
\usepackage{cite}
\usepackage{amsfonts}
\usepackage{amsmath}
\usepackage{pifont}
\usepackage{amssymb}
\usepackage[percent]{overpic}
\usepackage{xcolor}
\usepackage[export]{adjustbox}
\usepackage[T1]{fontenc}
\usepackage{hyperref}
\usepackage{tabu,multirow}
\usepackage[normalem]{ulem}

\newcommand{\cmark}{\ding{51}}%

\begin{document}
\title{Efficient Light Field Reconstruction \\ via Spatio-Angular Dense Network}
%
%
%

\author{Zexi~Hu,
        Henry~Wing~Fung~Yeung,
        Xiaoming~Chen,
        Yuk~Ying~Chung,~\IEEEmembership{Member,~IEEE}
        and~Haisheng~Li
\thanks{This work was supported by the Research Foundation for Advanced Talents of Beijing Technology and Business University (No. 19008021181), National Natural Science Foundation of China (No. 61877002), Beijing Natural Science Foundation and Fengtai Rail Transit Frontier Research Joint Fund (No. L191009) and Scientific Research Program of Beijing Municipal Education Commission (No. KZ202110011017).}
\thanks{Zexi Hu is with the School of Computer Science and Engineering, Beijing Technology and Business University, China and the School of Computer Science, University of Sydney, Australia. (Email: zexi.hu@sydney.edu.au)}
\thanks{Henry Wing Fung Yeung and Yuk Ying Chung are with the School of Computer Science, University of Sydney, Australia. (Email: henrywfyeung@gmail.com, vera.chung@sydney.edu.au)}
\thanks{Xiaoming Chen and Haisheng Li are with the School of Computer Science and Engineering, Beijing Technology and Business University, China. (Email: xiaoming.chen@btbu.edu.cn, lihsh@btbu.edu.cn)}
\thanks{Corresponding author: Xiaoming Chen.}
}

%
%

\markboth{Journal of \LaTeX\ Class Files,~Vol.~14, No.~8, August~2015}%
{Shell \MakeLowercase{\textit{et al.}}: Bare Demo of IEEEtran.cls for IEEE Journals}
%



\maketitle

\begin{abstract}
As an image sensing instrument, light field images can supply extra angular information compared with monocular images and have facilitated a wide range of measurement applications. Light field image capturing devices usually suffer from the inherent trade-off between the angular and spatial resolutions. To tackle this problem, several methods, such as light field reconstruction and light field super-resolution, have been proposed but leaving two problems unaddressed, namely domain asymmetry and efficient information flow. In this paper, we propose an end-to-end Spatio-Angular Dense Network (SADenseNet) for light field reconstruction with two novel components, namely correlation blocks and spatio-angular dense skip connections to address them. The former performs effective modeling of the correlation information in a way that conforms with the domain asymmetry. And the latter consists of three kinds of connections enhancing the information flow within two domains. Extensive experiments on both real-world and synthetic datasets have been conducted to demonstrate that the proposed SADenseNet's state-of-the-art performance at significantly reduced costs in memory and computation. The qualitative results show that the reconstructed light field images are sharp with correct details and can serve as pre-processing to improve the accuracy of related measurement applications. 
\end{abstract}


%
\IEEEpeerreviewmaketitle

\section{Introduction} \label{section:Introduction}
\IEEEPARstart{A}s an emerging image sensing instrument, light field (LF) cameras can capture a set of images from different perspectives. This feature offers advantages in vision-based measurement tasks. For example, researchers proposed using LF camera \cite{raghavendraLFFace_TIP2015, jiLFHOG_ICIP2016, sellahewaFace_TIM2010, fangFace_TIM2015} for more accurate face detection that is more robust to spoof attacks. Similar achievement in measurement accuracy has been seen in material recognition \cite{wangLFRecognition_ECCV2016, luLFRecognition_2019} where the subjects can't be easily distinguished in regular 2D images \cite{songEDRNet_TIM2020}, and salient object detection in complex scenarios \cite{shengLFSaliency_ICASSP2016, zhangLFNet_TIP2020}. The abundant information provided by LF instruments also facilitates depth measurement and 3D measurement with promising accuracy \cite{heberUshapeICCV2017, jeonDepthLightField2019, wangOcclusionawareDepthEstimation2015, chen2018accurate} compared with other types of image sensors such as stereo vision \cite{linStereoDepth_TIM2021} and structured light \cite{lilienblum3DStructuredLight_TIM2014}. The emergence of LF cameras, \textit{e.g.}, Raytrix R series \cite{website:Raytrix, heinze2016automated} and Lytro Illum \cite{website:Lytro}, has enabled LF instruments to apply to a wide range of consumer- and industrial-grade applications, such as measurement in industrial quality control \cite{heinze2016automated, deflectometry2021}. However, due to the limited capacity of micro-lens inside the sensor, these cameras inherently suffer from a low resolution and a trade-off between the angular and spatial resolutions.

For mitigating the trade-off, there are two major solutions, namely LF reconstruction and LF spatial super-resolution (LFSR). The former focuses on upsampling in the angular domain, \textit{i.e.} the number of sub-aperture images (SAI), while the latter aims at increasing the resolution of the spatial domain, \textit{i.e.} the spatial resolution of each SAI. This paper focuses on the former category to reconstruct a densely sampled LF from a sparsely sampled one. With the recent success of deep learning in image processing, learning-based methods are also introduced to LF reconstruction and have achieved superior performance \cite{kalantari_SIGGRAPHASIA2016, yeungSAS_ECCV2018}. However, these methods still suffer from problematic spatio-angular features extraction which limits their performance. It stems from two aspects, namely \textit{Domain Asymmetry} and \textit{Inefficient Information Flow}.

Regarding \textit{Domain Asymmetry}, it has been discovered in \cite{wuEPICNN_CVPR2017} that the information in the spatial and angular domains has distinct natures as the spatial domain contains the regular 2D image information while the angular domain encodes the disparity information between adjacent SAIs. It is also obvious that these two domains are spaces of enormously different sizes. Specifically, given a LF image $X \in \mathbb{R}^{U \times V \times W \times H}$, the spatial space $W \times H$ is usually much larger than the angular space $U \times V$, \textit{i.e.} $U \times V \ll W \times H$. This issue becomes more prominent in light field reconstruction where the high-frequency information in the angular domain is largely damaged when downsampling into the sparsely sampled LF image. In other words, the angular domain becomes further smaller and sparser. Therefore, it is unreasonable to process the two domains, which are asymmetrical in natures and volumes, in the same manner. Still, the existing methods manipulate the 4D LF data in a symmetrical manner. These methods majorly fall into three categories: 4D convolution filters \cite{yeungSAS_ECCV2018, yeungSAS_TIP2019}, epipolar-plane images (EPI) based methods \cite{wuEPICNN_CVPR2017, wuEPICNN_TPAMI2018, shinEPINET2018, heberUshapeBMVC2016, heberUshapeICCV2017, wangPseudo4DCNN2018} and pseudo-4D convolution filters \cite{wangLFRecognition_ECCV2016, yeungSAS_ECCV2018, yeungSAS_TIP2019}.

\begin{figure}
  \centering
  \begin{tabular}{@{}c@{}}
    \includegraphics[height=0.12\textheight]{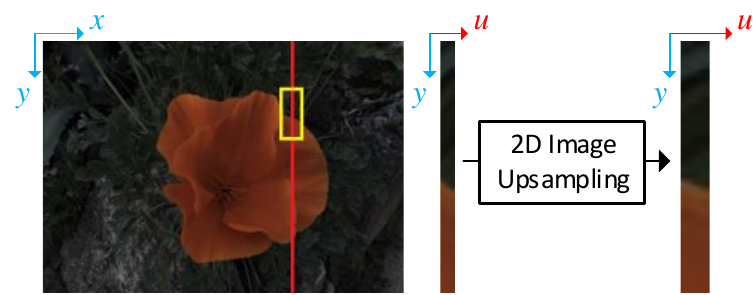}    \\
    \small(a)                                                        \\
    \includegraphics[height=0.13\textheight]{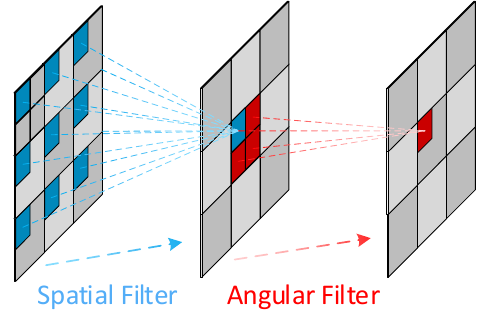} \\
    \small(b)                                                        \\
    \includegraphics[width=0.48\textwidth]{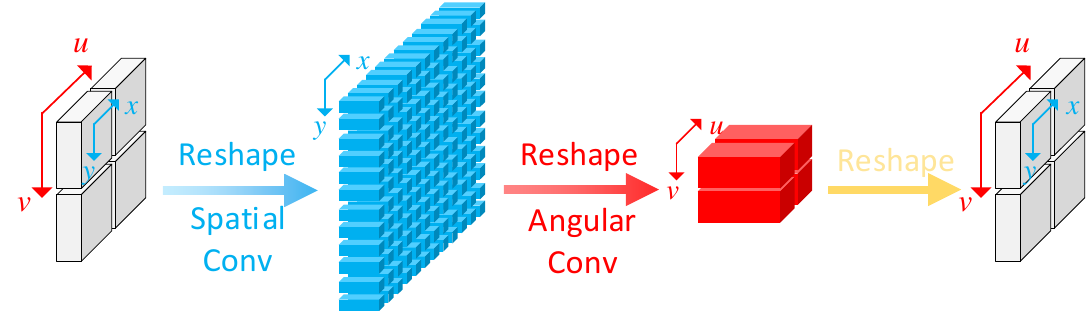}   \\
    \small(c)
  \end{tabular}

  \caption{The illustrations of the 4D convolution alternatives.
  (a) EPI, (b) Interleaved 2D convolution, (c) SAS convolution. They all process the spatio-angular information in a symmetrical way.
  }
  \label{fig:4D alternatives}
\end{figure}

A 4D convolution filter processes the LF image straightforwardly as it convolves the spatial and angular domain simultaneously. However, it is proved to be inefficient as it requires expensive computation and large memory consumption \cite{wangLFRecognition_ECCV2016, yeungSAS_ECCV2018}. As an alternative, the EPI-based methods have been proposed which perform independent super-resolution on EPI slices. An EPI slice can be obtained from the 4D LF by fixing 2 specific coordinates, with one lies on the spatial domain and the other on the angular domain. An illustration of this idea is depicted in Fig. \ref{fig:4D alternatives}(a). As EPI slices are essentially 2D images with patterns that reflect the correlation information in the LF image, these methods decompose the 4D LF reconstruction task into a series of 2D sub-tasks. Another alternative is pseudo-4D convolution filters which decompose a 4D convolution into a spatial and an angular convolution separately. Typical examples are the interleaved 2D angular and spatial convolutions to simulate 4D filters for material recognition \cite{wangLFRecognition_ECCV2016} and the spatio-angular separable (SAS) convolution for LF reconstruction and super-resolution \cite{yeungSAS_ECCV2018, yeungSAS_TIP2019}. Illustrations of EPI-based and pseudo-4D methods are demonstrated in Fig. \ref{fig:4D alternatives} (b) and (c) correspondingly.

Even though the aforementioned three options have achieved remarkable progress in extracting spatio-angular features, the vast majority of them treat the spatial and angular dimensions symmetrically against the domain asymmetry. One of the very few exceptions is \cite{wuEPICNN_CVPR2017} where the issue was addressed by its "blur-restoration-deblur" framework which downsamples the spatial domain for conforming with the asymmetry at the expense of losing spatial high-frequency information rendering the method sub-optimal. In this paper, we argue that the operations performed in the spatial and angular domains should be of distinct natures accordingly. The convolution operation performed in the spatial domain is modeling the local spatial features such as edges and corners, whereas the one performed in the angular domain is modeling the disparities information. Therefore, we have hypothesized that the symmetrical pattern hampers the spatio-angular feature representation and we proposed \textit{correlation blocks} which comprise an uneven number of spatial and angular convolutions to model correlation information of LF images asymmetrically. Unlike \cite{wuEPICNN_CVPR2017}, our approach doesn't involve any down-sampling or other operations that would explicitly abandon any valuable information in any domain but fully convolves on both domains. We elaborate this component in Section \ref{section:Correlation Block} and validate our hypothesis in the ablation study in Section \ref{section:Ablation Study Asymmetry}.

As for \textit{Inefficient Information Flow}, most computer vision methods have suffer from it as they opted to increase network depth \cite{simonyanVGG_2015, kim_accurate_2016} to extract deeper features. Obviously, its downside is a huge number of trainable parameters and high computation cost. The other defect is the gradient-vanishing problem \cite{huangDenseNet_CVPR2017}, especially in shallow layers, as networks go deeper, resulting in training difficulties. These negative effects can be amplified in the realm of LF images due to its 4D high volume \cite{wangLFRecognition_ECCV2016, yeungSAS_ECCV2018}, leading to model inefficiency when training and testing. For example, in \cite{yeungSAS_ECCV2018}, Yeung \textit{et al.} have proposed a network that outperforms Kalantari \textit{et al.} \cite{kalantari_SIGGRAPHASIA2016} by 0.3 db PSNR with 4 SAS convolutions. Yet, the performance gain is diminishing as more SAS convolutions are added, with a further 0.3 db and 0.6 db improvement at 8 and 16 SAS convolutions, which requires 1.5 and 2 times of model size respectively. Although it has achieved state-of-the-art performance, the 16-SAS-convolution network contains around 1.5 million parameters and takes more than 8 days to fully converge, rendering the model difficult to train and impractical to apply to memory-constrained devices.

Inspired by the success of dense skip connections which have been widely studied and exploited in image processing \cite{huangDenseNet_CVPR2017, taiMemNet_ICCV2017, tongSRDenseNet_ICCV2017, harisDBPN_CVPR2018, liSRFBN_CVPR2019}, we propose to use dense skip connections to enhance the information flow. Contrary to the previous dense networks, we propose spatio-angular dense skip connections specially designed for LF images which consist of three kinds of connections, namely angular, spatial, and image skip connections, to supply distinct types of information and reinforce the spatio-angular feature representation.

Lastly, with these two components, correlation blocks and spatio-angular dense skip connections, we propose a simple yet efficient end-to-end Spatio-Angular Dense Network (SADenseNet). Extensive experiments are conducted on both real-world and synthetic datasets to demonstrate SADenseNet's superior performance of LF reconstruction at substantially lower computational and memory costs compared with state-of-the-art methods. A series of ablation studies are also presented to verify the proposed components' effectiveness, and we also perform SADenseNet as pre-processing that yields a positive effect for depth estimation to prove its potential for more accurate LF measurement application.

The major contributions of this paper are summarized in the following five aspects:
\begin{itemize}
    \item Correlation blocks are proposed to model correlation information based on the study of domain asymmetry.
    \item Spatio-angular dense skip connections are proposed to enhance the information flow within spatial and angular domains.
    \item With the proposed components, we design a simple yet efficient end-to-end network, Spatio-Angular Dense Network (SADenseNet), for light field reconstruction.
    \item SADenseNet is evaluated by extensive experiments on both real-world and synthetic datasets to verify its superior performance and efficiency in computation and memory usage compared with previous state-of-the-art methods.
    \item An experiment on depth estimation is conducted to prove SADenseNet can improve the accuracy of LF's measurement applications.
\end{itemize}

\section{Related Works}
\subsection{Light Field Image Processing}
To process LF images, early methods \cite{pujadesBayesianViewSynthesis2014, zhang2015light, wanner2014variational, zhang_light_2015} explicitly estimate the disparity or depth and then obtain the reconstructed SAIs by warping the input SAIs and blending. Various depth estimation methods are proposed, \textit{e.g.}, phase-based estimation \cite{zhang_light_2015} and EPI-based estimation \cite{wanner2014variational}. Different blending methods are also introduced, \textit{e.g.}, soft-blending \cite{penner_soft_2017} and learning-based blending \cite{zhengCrossNet_ECCV2018}.
These methods are overly dependent on the quality of the disparity or depth maps which the existence of noise in them causes undesirable artifacts. Such approaches are also disadvantaged when handling occluded areas as the warping process has no information about the invisible parts.

In recent years, significant progress in computer vision has been achieved by the deep learning-based methods \cite{simonyanVGG_2015, pengLargeKernelGCN_CVPR2017, caballeroVESPCN_CVPR2017, tongSRDenseNet_ICCV2017}, and this technique has been applied to LF reconstruction.
The first deep learning-based LF reconstruction method has been proposed by Kalantari \textit{et al.} \cite{kalantari_SIGGRAPHASIA2016} which designed a disparity network to estimate the disparity and reconstruct the intermediate SAIs by warping the input SAIs with the calculated disparity. After that, the input and intermediate SAIs are fed into a color network to obtain the refined SAIs. This method has achieved outstanding performance with the deep features extracted by the deep neural networks. However, the reconstruction quality is limited due to the artifacts brought by the explicit disparity estimation and warping process. Moreover, each SAI needs to be individually reconstructed, which leads to duplicated calculations. Similarly,  Zhou \textit{et al.} \cite{ZhouNoisyLFRecon_SS2021} proposed an encoder-decoder network to estimate the disparity to synthesize views by warping. With the help of its modified ResNet-50 to extract expressive representation and three sub-networks for disparity estimation, noise filtering, and view rendering respectively, robustness is gained against input noise.

Wang \textit{et al.} proposed an EPI-based method to decompose the 4D task into 3D sub-tasks and train two networks separately for vertical and horizontal reconstruction \cite{wangPseudo4DCNN2018}. Heber \textit{et al.} \cite{heberUshapeBMVC2016, heberUshapeICCV2017} designed a U-shaped network to extract shape information by EPIs while Shin \textit{et al.} \cite{shinEPINET2018} proposed a EPI-based method to compute disparity maps from LF with a four-branch network. A common drawback of EPI-based methods is they ignore either one \cite{wangPseudo4DCNN2018, shinEPINET2018, heberUshapeBMVC2016} or two dimensions \cite{wuEPICNN_CVPR2017, heberUshapeICCV2017} of the 4D LF in their optimization and cannot be jointly optimized in all dimensions. Therefore, they cannot process the LF data in a global scope.

In \cite{yeungSAS_ECCV2018}, Yeung \textit{et al.} proposed an end-to-end network that removes explicit disparity manipulation and reconstructs SAIs at a single forward propagation. This method has achieved groundbreaking performance with its SAS convolution that enables joint optimization in all dimensions. However, it has endured the same defect of stacking convolutional layers for obtaining deeper feature representation as other computer vision tasks \cite{simonyanVGG_2015, szegedyGoogLeNet_CVPR2015}. Different from decomposing 4D convolution \cite{yeungSAS_ECCV2018}, Meng \textit{et al.} \cite{mengHDDRNet_TPAMI2019} pursued to fully exploit high-dimensional LF information and proposed aperture group batch normalization to ease the training of 4D convolutional network. In addition to pixel-wise loss function in the spatial space, they proposed an angular loss to optimize the error in the EPI space to preserve the correlation information of adjacent viewpoints. However, suffering from heavy 4D convolution operations, it was still running at a large cost of computation and memory, and the high-dimensional information was yet to be fully exploited as the performance gain was not proportional to its costs. Besides pure learning-based methods, Chandramouli \textit{et al.} \cite{ChandramouliGenerativeLF_TPAMI2020} proposed a generative model to tackle the issue that learning-based models are confined to the observation model they have been trained on. Despite its limited performance, the fashion is inspiring with possibilities of LF models with generalization abilities.

There are some other methods \cite{GuoLFCA_ECCV2020, InagakiCodedAperture_ECCV2018} that perform LF reconstruction but in a compressive manner. Their compression performance is proved to be promising, but they are essentially a different task from regular LF reconstruction as the full LF image is involved to be encoded into a small volume of information instead of a limited set of SAIs. Although the above methods and the regular LF reconstruction are not working in exactly the same application scenarios, it would be interesting to see where their performance ceilings are located.

In this paper, our proposed SADenseNet has a similar pipeline as \cite{yeungSAS_ECCV2018} that contains two parts, namely the feature extraction and the SAI reconstruction. However, we ditch the refinement work as SADenseNet's feature extraction is so strong with the help of correlation blocks and spatio-angular dense skip connections that post-processing components are unnecessary. Rather than inefficient explicit warping process \cite{kalantari_SIGGRAPHASIA2016} and costly 4D convolution operations \cite{mengHDDRNet_TPAMI2019}, we pursue the potential of decomposed convolution operations and information flow enhancement.

\begin{figure*}[ht]
    \centering
    \includegraphics[width=0.99\textwidth]{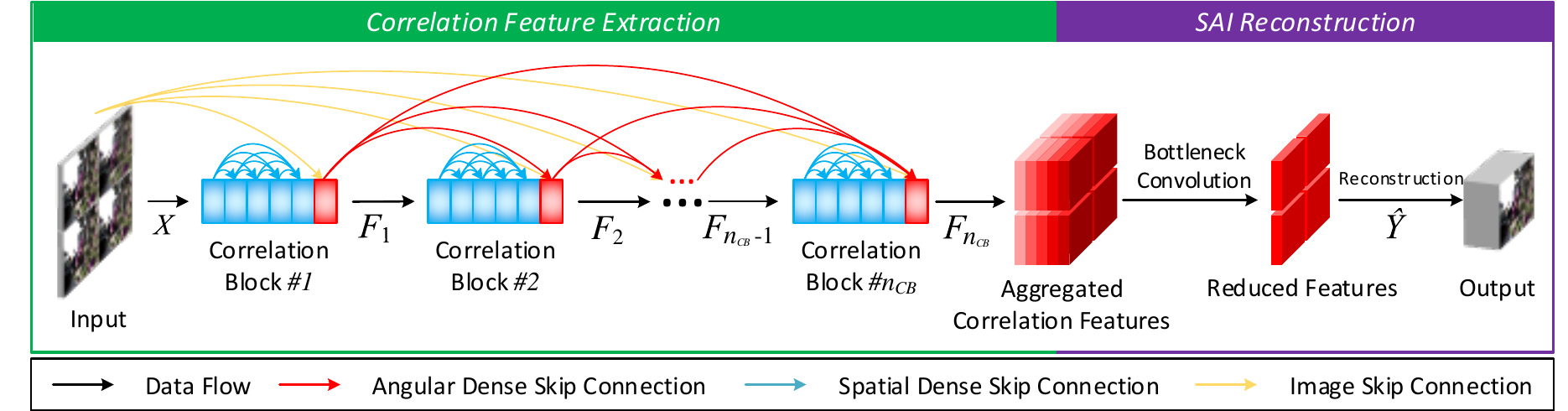}
    \caption{The illustration of SADenseNet, which comprises two processes, namely correlation feature extraction and SAI reconstruction. The black arrows indicate the regular data flow and the colored arrows indicate dense skip connections. Red and blue arrows represent angular and spatial dense skip connections, and yellow ones represent image skip connections.}
    \label{fig:SADenseNet}
\end{figure*}

\subsection{Feature Representation}
In order to acquire powerful feature representation, stacking convolutional layers is the most straightforward solution since more parameters lead to higher non-linearity and more complicated mapping functions. However, heavier computation costs and higher risks of over-fitting are also incurred. To this end, some studies have discovered more efficient network structures.

The residual connection is one of the widely used structures firstly proposed by He \textit{et al.} \cite{heResNet_CVPR2016} to force layers to learn residue of a mapping function instead of the mapping function itself. Such a technique eases the optimization and improves the feature representation, and can be applied to other computer vision tasks such as single image super-resolution (SISR) \cite{kim_accurate_2016,tai_image_2017,zhangRCAN_ECCV2018} and visual tracking \cite{songCREST_ICCV2017, wang_learning_2018}.

On the other hand, dense skip connections have been proposed to mitigate the gradient-vanishing problem while aggregating hierarchical features of low- and high-frequencies \cite{huangDenseNet_CVPR2017}. Tai \textit{et al.} \cite{taiMemNet_ICCV2017} proposed a dense connected network to aggregate short-term and long-term memories for image restoration. In SISR, Tong \textit{et al.} designed local and global dense skip connections to aid the feature extraction \cite{tongSRDenseNet_ICCV2017}. Haris \textit{et al.} \cite{harisDBPN_CVPR2018} designed a projection unit to learn down-sampling as well as up-sampling. Li \textit{et al.} \cite{liSRFBN_CVPR2019} introduced feedback units containing projection groups to form a recurrent neural network (RNN) to enforce a curriculum learning strategy. The basic units of the latter two methods \cite{harisDBPN_CVPR2018, liSRFBN_CVPR2019} are reinforced with dense skip connections to gain richer feature representation. Among these, \cite{taiMemNet_ICCV2017} and \cite{tongSRDenseNet_ICCV2017} have similar architectures with our SADenseNet as they perform dual dense skip connections. Nevertheless, the critical difference is that their duality is for extracting homogeneous features while ours are for spatio-angular features from the two distinct domains.

\subsection{Video-related Tasks}
Video-related tasks share similarities with LF tasks as they are both extracting cross-domain representation and processing multiple images or frames that are highly correlated to each other. A powerful spatio-temporal feature representation is a key to success as well. In terms of decomposition of high dimensions, similar methods can be found in pseudo-3D-based P3D \cite{qiuP3D_ICCV2017, liP3DSR_CVPR2019} and R2D \cite{tranR2D_CVPR2018}. Different from the pseudo-4D convolution in LF, the pseudo-3D convolutions decompose a 3D convolution into two 2D convolutions operating in the spatial and temporal spaces separately. Convolution decomposition has been proved to achieve better performance than the original 3D convolution achieved by the extra non-linearity and the reduced risk of over-fitting.

\section{Proposed Method}
\subsection{Overview}
Let us consider $\mathcal{F}(X)$ as the LF reconstruction function that maps the input SAIs $X$ to the reconstructed SAIs $\hat{Y}$, hence
\begin{equation}\label{eq:formula}
\begin{split}
    \hat{Y} = \mathcal{F}(X) & \quad X \in \mathbb{R}^{U \times V \times W \times H} \\
    & \quad \hat{Y} \in \mathbb{R}^{\mathcal{N} \times W \times H}
\end{split}
\end{equation}

\noindent where $(U, V)$ of $X$ and ${\mathcal{N}}$ of $\hat{Y}$ denote the input and output angular resolution respectively, \textit{i.e.} the number of SAIs, while $(W, H)$ is the spatial resolution, \textit{i.e.} the resolution of every single SAI. The input SAIs $X$, which is a sparsely sampled LF image, and the reconstructed SAIs $\hat{Y}$ can finally form a densely sampled LF image ${\hat{I} \in \mathbb{R}^{\overline{U} \times \overline{V} \times W \times H}}$. Therefore, 

\begin{equation*}
\overline{U} \times \overline{V} = U \times V + \mathcal{N}.
\end{equation*}

The objective of this paper is to train an end-to-end network to learn $\mathcal{F}(X)$ in Eq. \ref{eq:formula} for estimating $\hat{Y}$ that approximates the ground truth $Y$. The proposed SADenseNet comprises two components, correlation feature extraction, and SAI reconstruction, and an illustration is given in Fig. \ref{fig:SADenseNet}. The input SAIs $X$ are fed into correlation feature extraction containing a series of correlation blocks, which will be described in Section \ref{section:Correlation Block}. In addition to the correlation blocks, spatio-angular dense skip connections are exploited through the inter- and intra-correlation-blocks to enhance the information flow, which will be elaborated in Section \ref{section:Spatio-Angular Dense Skip Connection}. With the extracted correlation features, SAI reconstruction will finally reconstruct SAIs $\hat{Y}$, which will be explained in Section \ref{section:SAI Reconstruction}.

\begin{figure*}
    \centering
    \includegraphics[width=0.98\textwidth]{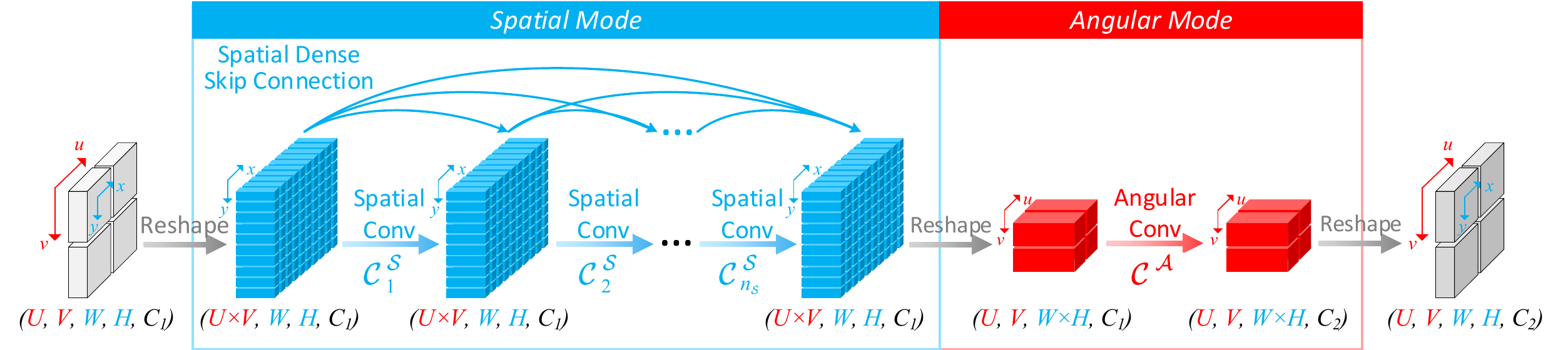}
    \caption{The illustration of a correlation block. Blue tensors and arrows indicate tensors and operations under the spatial mode. Red counterparts indicate the angular ones likewise.}
    \label{fig:CorrelationBlock}
\end{figure*}

\subsection{Correlation Blocks} \label{section:Correlation Block}

For addressing the aforementioned domain asymmetry problem in Section \ref{section:Introduction}, we propose correlation blocks that process the spatial and angular information asymmetrically. The structure is demonstrated in Fig. \ref{fig:CorrelationBlock}.

The component is based on the SAS module proposed in \cite{yeungSAS_ECCV2018, yeungSAS_TIP2019}. However, motivated by the fact that the spatial space is usually substantially larger compared to the angular space, \textit{i.e.} $U \times V \ll W \times H$, we presume that more spatial operations are needed than angular operations. Therefore, in each block, the 4D LF tensor of size $(U, V, W, H, C_1)$, where $C_1$ is the number of input channels, is firstly reshaped into the spatial mode of size $(U \times V, W, H, C_1)$ to flatten the angular dimensions. Subsequently, a series of spatial convolutional layers, denoted as $[\mathcal{C}^{\mathcal{S}}_1, \mathcal{C}^{\mathcal{S}}_2, \dotsb, \mathcal{C}^{\mathcal{S}}_{n_{\mathcal{S}}}]$, are operated on the reshaped tensor of kernel size $(1, k_W, k_H, C_1, C_1)$ to extract the spatial features, where $n_{\mathcal{S}}$ is the number of spatial convolutional layers\footnote{In this paper, the last two values of the size of 3D convolution kernels denote the input and output channels correspondingly.}. Likewise, the output of spatial convolution series is reshaped into the angular mode of size $(U, V, W \times H, C_1)$ and convolved by an angular convolutional layer of kernel size $(k_U, k_V, 1, C_1, C_2)$ to extract angular features that model the correlation information, where $C_2$ is the number of output channels. Lastly, the tensor is reshaped back into the original shape with new channels $(U, V, W, H, C_2)$ for the upcoming operation. Hence, a correlation block can be formulated as
\begin{gather} 
\mathcal{H}(x) = \mathcal{A}(\mathcal{S}(x)) \\ 
\mathcal{A}(x) = \mathcal{C}^{\mathcal{A}}(x) \\
\mathcal{S}(x) = \mathcal{C}^{\mathcal{S}}_{n_{\mathcal{S}}}(\mathcal{C}^{\mathcal{S}}_{n_{\mathcal{S}}-1}(\dotsb(\mathcal{C}^{\mathcal{S}}_{2}(\mathcal{C}^{\mathcal{S}}_{1}(x)))\dotsb) \label{eq:naiveConvS}
\end{gather}

\noindent where $\mathcal{S}$ and $\mathcal{A}$ denote the spatial and angular feature extraction while $\mathcal{C}^{\mathcal{S}}(x)$ and $\mathcal{C}^{\mathcal{A}}(x)$ denote the corresponding convolution functions.

Such an asymmetrical processing method suits LF image data compared with the existing 4D convolution alternatives because the spatial information is sufficiently convolved to extract informative features with an enlarged receptive field in the spatial domain before the subsequent angular convolution.

In order to acquire deeper correlation features, a series of correlation blocks are employed consecutively as

\begin{equation} \label{eq:naiveCB}
F_i = 
\begin{cases}
      \mathcal{H}(F_{i-1}), & \text{if}\ i>1, \\
      \mathcal{H}(X), & \text{if}\ i=1.
\end{cases}
\end{equation}
where $F_i$ denotes the output tensor of $i$-th correlation block.

The angular convolutions of correlation blocks are operated on the coarse-to-fine spatial features, thus they can extract correlation features hierarchically. The correlation extraction can be regarded as implicit disparity modeling but without explicit dependency on the accuracy of disparity. We denote the number of correlation blocks as $n_{CB}$ and set both $(k_U, k_V)$ and $(k_W, k_H)$ to $(3,3)$ across all the correlation blocks. Finally, the output of the last correlation block $F_{n_{CB}}$ will be fed to the following SAI reconstruction process.

\subsection{Spatio-Angular Dense Skip Connections} \label{section:Spatio-Angular Dense Skip Connection}
For further enhancing the feature representation extraction process, we propose to improve the information flow with spatio-angular dense skip connections consisting of three types of connections, namely spatial and angular dense skip connections, and image skip connections.

The spatial dense skip connections are intra-correlation-block connections that concatenate the output of shallow spatial convolutional layers to the output of the subsequent layers within a correlation block, \textit{i.e.} the spatial convolutional layers will receive the features from the preceding layers. Consequently, the spatial convolution function in Eq. \ref{eq:naiveConvS} can be revised as
\begin{gather}
\mathcal{S}(x) = \tilde{\mathcal{C}}^{\mathcal{S}}_{n_S}(x) \\
\tilde{\mathcal{C}^{\mathcal{S}}_{i}}(x) = [\mathcal{C}^{\mathcal{S}}_{i}(\tilde{\mathcal{C}}^{\mathcal{S}}_{i-1}(x)), \tilde{\mathcal{C}}^{\mathcal{S}}_{i-1}(x), \dotsb, \tilde{\mathcal{C}}^{\mathcal{S}}_2(x), \tilde{\mathcal{C}}^{\mathcal{S}}_1(x)]
\end{gather}
where $\tilde{\mathcal{C}^{\mathcal{S}}_{i}}(x)$ represents the $i$-th densely connected spatial convolutional layer and $[\cdot]$ indicates concatenation. The spatial dense skip connections are demonstrated in Fig. \ref{fig:SADenseNet} and \ref{fig:CorrelationBlock} as blue arrows connecting blue tensors under the spatial mode. 

The densely connected structure comes with three critical benefits. Firstly, the feature tensors are reinforced with both shallow and deep information, forming hierarchical representations. This is contrary to the previous methods \cite{yeungSAS_ECCV2018} where the shallow features are absent in high-level processing. Secondly, it facilitates the optimization of the network, especially for the early layers, since every layer will access the gradients directly in back-propagation and the gradient vanishment problem is also greatly alleviated. Last but not least, the high trainable parameter utilization leads to model efficiency and prevents over-fitting.

While the spatial dense skip connections strengthen the spatial representation within individual correlation blocks, we explore to bring this benefit to the angular domain by adopting angular dense skip connections that concatenate the output of the preceding correlation blocks $F_i$ to the latter ones. As angular convolutions are connected with their counterparts in the previous correlation blocks, the angular dense skip connections play a role as inter-correlation-block information flow as depicted in Fig. \ref{fig:SADenseNet} as the red arrows.

Furthermore, inspired by the practice of appending raw input to the intermediate layers in video processing \cite{caballeroVESPCN_CVPR2017} and optical flow \cite{ilgFlowNet2_CVPR2017}, we introduce image skip connections to provide input $X$ as primitive features for all the blocks. This design is very similar to \cite{caballeroVESPCN_CVPR2017, ilgFlowNet2_CVPR2017} as these two methods implicitly estimate the motion between frames and inject raw image information into the intermediate motion feature for further refinement, while in our case, the correlation blocks implicitly model the LF disparity and the raw image offers complementary information for the angular convolutions.
The image skip connections are depicted as yellow arrows connecting the input LF and the correlation blocks in Fig. \ref{fig:SADenseNet}.

Accordingly, the function of correlation blocks in Eq. \ref{eq:naiveCB} are revised into
\begin{equation}
F_i = 
\begin{cases}
        [\mathcal{H}(F_{i-1}), F_{i-1}, \dotsb, F_{1}, X], & \text{if}\ i>1, \\
        \mathcal{H}(X), & \text{if}\ i=1
\end{cases}
\end{equation}
where $[\cdot]$ indicates concatenation. The output of the last correlation block $F_{n_{CB}}$ aggregates the feature maps of all the correlation blocks to form hierarchical spatio-angular feature representation. It will be fed to the following SAI reconstruction process.

In \cite{huangDenseNet_CVPR2017}, the number of feature maps increased by dense skip connections is referred to as the \textit{growth rate}. For simplicity, we keep the growth rates in both domain consistent across the network. Therefore,
\begin{equation*}
r_\mathcal{S} = r_{CB} = C_1 = C_2   
\end{equation*}
where $r_\mathcal{S}$ and $r_{CB}$ are growth rates of the spatial and angular dense connections respectively. Consequently, the number of feature maps of the last correlation block $F_{n_{CB}}$ and the last spatial convolution $\tilde{\mathcal{C}}^{\mathcal{S}}_{n_S}(x)$ in each correlation block will grow to $r_{CB} \times n_{{CB}}$ and $r_\mathcal{S} \times n_{\mathcal{S}}$ respectively. 
In the experiments of Section \ref{section:Experiments}, the growth rates are set to 32, which are only half of the feature map number used in \cite{yeungSAS_ECCV2018}, by virtue of the efficiency of densely connected architecture.

\subsection{SAI Reconstruction} \label{section:SAI Reconstruction}
Before reconstruction using the extracted hierarchical correlation features, we follow the practice in \cite{tongSRDenseNet_ICCV2017} to employ a bottleneck layer to aggregate the hierarchical features and reduce the number of the feature maps to 96. Then, the reduced features are reshaped into the angular mode, zero-padded, and fed into an angular convolutional layer with a kernel size of $(U, V, 1, 96, \mathcal{N})$. It shrinks the angular resolution from $(U \times V)$ to $(1 \times 1)$ while producing $\mathcal{N}$ channels, which correspond to the $\mathcal{N}$ reconstructed SAIs. In Fig. \ref{fig:SADenseNet}, as $(U \times V)=(2 \times 2)$ and $\mathcal{N}=60$, the kernel size of the angular convolution is $(2, 2, 1, 96, 60)$.
\begin{table*}[ht]
\centering
\caption{Overall comparison with the state-of-the-art on the real-world datasets. The table is divided into two parts: 1. The 2nd to 4th columns are PSNR and SSIM on four datasets. The number of image samples in a particular dataset is shown in parentheses. 2. The last two columns display the number of parameters and running speed. Bold scores indicate the best results. The last row is the differences between SADenseNet and Yeung et al. \cite{yeungSAS_ECCV2018}.}
\label{tab:OverallPerformance}
\begin{tabu}{|l|[2pt]c|c|c|c|[2pt]c|c|}
\hline
Method                                                          & 30 Scenes (30)           & EPFL (118)            & Occlusions (43)       & Reflective (31)        & \# Parameters     & Speed/s  \\ \hline\hline
Kalantari \textit{et al.} \cite{kalantari_SIGGRAPHASIA2016}     & 38.31/0.9755             & 38.76/0.9586          & 31.82/0.8973          & 35.91/0.9415           & 1,644,204           & 721.05               \\ \hline
HDDRNet \cite{mengHDDRNet_TPAMI2019}                            & 37.52/0.9664             & 38.58/0.9547          & 32.36/0.9071          & 36.32/0.9443           & 16,558,848           & 1434.65               \\ \hline
NoisyLFRecon \cite{ZhouNoisyLFRecon_SS2021}                     & 38.90/0.9776             & 39.01/0.9639          & 32.12/0.9070          & 36.36/0.9455           & 20,277,551           & 32.92               \\ \hline
Yeung \textit{et al.} \cite{yeungSAS_ECCV2018}                  & 39.16/0.9782             & 39.53/0.9641          & 32.66/0.9073          & 36.44/0.9458           & 1,498,752           & 38.05               \\ \hline
SADenseNet(Ours)                                                & \textbf{40.31/0.9836}    & \textbf{40.54/0.9706} & \textbf{33.76/0.9269} & \textbf{37.15/0.9521}  & \textbf{1,134,140}  & \textbf{12.82}  \\ \hline\hline    
Difference                                                      & +1.15/+0.0054            & +1.01/+0.0065         & +1.10/+0.0196         & +0.71/+0.0063          & 75.67\%               & $\approx 3 \times$ faster  \\ \hline
\end{tabu}
\end{table*}

\begin{table}[t]
\centering
\tabcolsep=0.15cm
\caption{Performance comparison with the state-of-the-art on the synthetic HCI dataset \cite{wannerHCI2013}. PSNR and SSIM are given. Bold scores indicate the best results. The differences between SADenseNet and Yeung et al. \cite{yeungSAS_ECCV2018} are presented in the last row.} 
\label{tab:HCIPerformance}
\begin{tabu}{|l|[2pt]c|c|c|}
\hline
Method                                                      & \textit{Buddha}           & \textit{Mona}                 & Average  \\ \hline\hline
Kalantari \textit{et al.} \cite{kalantari_SIGGRAPHASIA2016} & 42.73/0.9844              & 42.42/0.9858                  & 42.58/0.9851 \\ \hline
Wu \textit{et al.} \cite{wuEPICNN_TPAMI2018}                & 43.20/\textbf{0.9963}     & 44.37/\textbf{0.9977}         & 43.79/\textbf{0.9981} \\ \hline
Yeung \textit{et al.} \cite{yeungSAS_ECCV2018}              & 43.77/0.9872              & 45.67/0.9920                  & 44.72/0.9896 \\ \hline
SADenseNet(Ours)                                            & \textbf{45.82}/0.9921     & \textbf{46.84}/0.9932         & \textbf{46.33}/0.9927 \\ \hline\hline    
Difference                                                  & +2.05/+0.0049             & +1.17/+0.0012                 & +1.61/+0.0031 \\ \hline
\end{tabu}
\end{table}

\section{Experiments} \label{section:Experiments}

\subsection{Implementation and Evaluation Details}
The proposed SADenseNet is implemented using the deep learning library Keras \cite{keras} with Tensorflow \cite{tensorflow} backend. The network is trained and tested on a PC with an Intel Core i7-6700K 8-core 4.00GHz CPU, an Nvidia GTX 1080 Ti GPU, and 32GB RAM.
The source code and trained models are publicly available at \href{https://huzexi.github.io}{https://huzexi.github.io}.

In regard to training, to conduct a fair comparison, we follow the protocol of \cite{yeungSAS_ECCV2018} to train the network with the training set proposed by Kalantari \textit{et al.} \cite{kalantari_SIGGRAPHASIA2016}, which contains 100 samples. The mini-batch size is set to 2 with a spatial size of $(128 \times 128)$. The training process is iterated with an Adam optimizer \cite{kingmaAdam_2014} and the learning rate is set to $1e-4$. We follow the conventional practice to process the luminance channel of the YCbCr color space. The other two channels, namely Cb and Cr, are acquired by up-sampling the angular resolution using bicubic interpolation. During training, data augmentation is applied to improve the generalization of the network. We follow the strategies in \cite{shinEPINET2018} to randomly flip and rotate the spatial and angular dimensions simultaneously. As a result, the training data can be reused 8 times.

$n_{CB}$ and $n_\mathcal{S}$ are set to 6 and 5 correspondingly as the default values, and the network is trained by minimizing the Mean Squared Error (MSE) loss function as follows:
\begin{equation}\label{eq:loss}
\min_{\hat{Y}}{ \sum_{n}^{\mathcal{N}}\sum_{x}^{W}\sum_{y}^{H}     (\hat{Y}(n, x, y) - Y(n, x, y)) ^2 }
\end{equation}
where $n$ iterates over the reconstructed SAIs and $(x,y)$ indicates a spatial location.

\subsection{Comparison with State-of-the-art Methods on Real-world Images} \label{section: ExpRealWorld}
Firstly, we compare SADenseNet's performance on real-world images with the state-of-the-art methods: Kalantari \textit{et al.} \cite{kalantari_SIGGRAPHASIA2016}, HDDRNet \cite{mengHDDRNet_TPAMI2019}, NoisyLFRecon \cite{ZhouNoisyLFRecon_SS2021} and Yeung \textit{et al.} \cite{yeungSAS_ECCV2018}.

Experiments are conducted on four real-world LF datasets, namely 30 Scenes \cite{kalantari_SIGGRAPHASIA2016}, EPFL \cite{rerabekEPFL2016}, Occlusions \cite{StanfordLytro} and Reflective \cite{StanfordLytro} which are captured by Lytro Illum cameras \cite{website:Lytro}. These four datasets contain 30, 118, 43, and 31 LF images correspondingly. The LF images that have appeared in the 100-sample training set are removed for fairness. In each LF image, the angular resolution is $(14 \times 14)$ and the spatial resolution is $(376 \times 541)$. During the evaluation, only the central $(8 \times 8)$ SAIs are adopted as the rest are dark and noisy, 22 pixels of the four borders in the spatial resolution are also shaved as in \cite{kalantari_SIGGRAPHASIA2016}. The reconstruction is performed from $(2 \times 2)$ SAIs to $(8 \times 8)$ SAIs. The reconstruction quality is measured by Peak Signal-to-Noise Ratio (PSNR) and Structural Similarity Index (SSIM) in RGB color space.

The results are shown in Table \ref{tab:OverallPerformance}. It can be observed that SADenseNet has outperformed the other methods observably. Compared with the best of them, Yeung \textit{et al.} \cite{yeungSAS_ECCV2018}, SADenseNet outperforms by more than 1.00 dB PSNR in all datasets except the Reflective dataset where our proposed approach has a reconstruction advantage of 0.71 dB PSNR.

\begin{figure*}[t]
    \centering
    \tabcolsep=0.04cm
    \renewcommand{\arraystretch}{0.6}
    \begin{tabular}{cccccccccc}
        \multicolumn{2}{c}{Ground truth}
        & \multicolumn{2}{c}{HDDRNet}
        & \multicolumn{2}{c}{NoisyLFRecon}
        & \multicolumn{2}{c}{Yeung \textit{et al.} }
        & \multicolumn{2}{c}{SADenseNet (Ours)}
        \\

        \multicolumn{2}{c}{\includegraphics[width=0.192\textwidth]{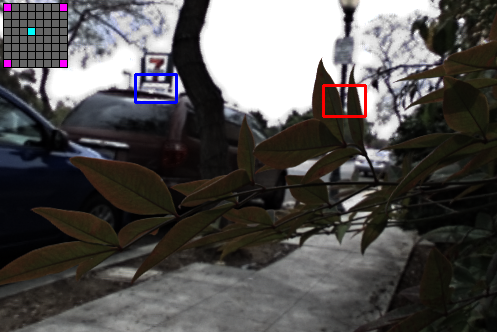}}
        & \multicolumn{2}{c}{\includegraphics[width=0.192\textwidth]{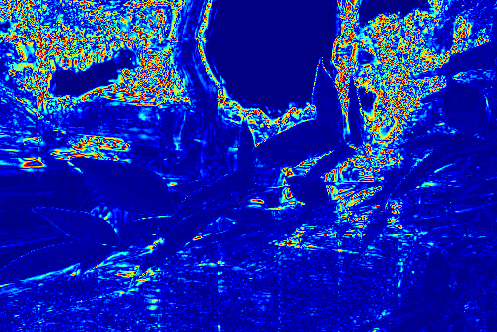}}
        & \multicolumn{2}{c}{\includegraphics[width=0.192\textwidth]{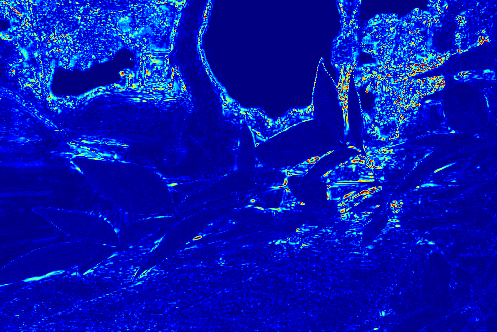}}
        & \multicolumn{2}{c}{\includegraphics[width=0.192\textwidth]{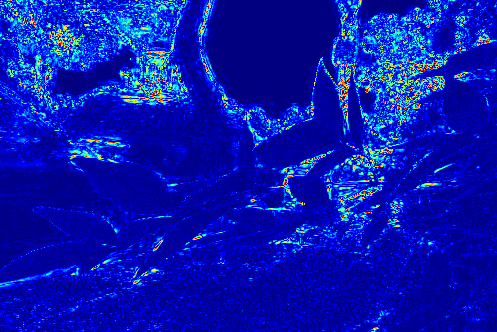}}
        & \multicolumn{2}{c}{\includegraphics[width=0.192\textwidth]{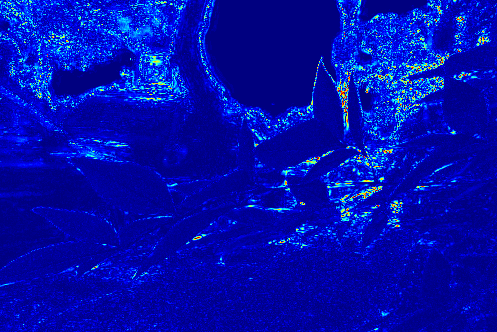}}
        \\

        \includegraphics[width=0.09\textwidth, height=0.052\textwidth, cfbox=red 1pt 0pt]{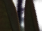}
        & \includegraphics[width=0.09\textwidth, height=0.052\textwidth, cfbox=blue 1pt 0pt]{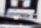}
        & \includegraphics[width=0.09\textwidth, height=0.052\textwidth, cfbox=red 1pt 0pt]{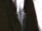}
        & \includegraphics[width=0.09\textwidth, height=0.052\textwidth, cfbox=blue 1pt 0pt]{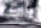}
        & \includegraphics[width=0.09\textwidth, height=0.052\textwidth, cfbox=red 1pt 0pt]{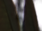}
        & \includegraphics[width=0.09\textwidth, height=0.052\textwidth, cfbox=blue 1pt 0pt]{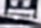}
        & \includegraphics[width=0.09\textwidth, height=0.052\textwidth, cfbox=red 1pt 0pt]{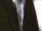}
        & \includegraphics[width=0.09\textwidth, height=0.052\textwidth, cfbox=blue 1pt 0pt]{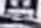}
        & \includegraphics[width=0.09\textwidth, height=0.052\textwidth, cfbox=red 1pt 0pt]{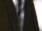}
        & \includegraphics[width=0.09\textwidth, height=0.052\textwidth, cfbox=blue 1pt 0pt]{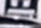}
        \\

        \multicolumn{2}{c}{\textit{IMG\_1528}}
        & \multicolumn{2}{c}{25.35/0.8926}
        & \multicolumn{2}{c}{31.15/0.9540}
        & \multicolumn{2}{c}{31.30/0.9543}
        & \multicolumn{2}{c}{34.18/0.9715}
        \\
        \vspace{-0.15cm} \\
        
        \multicolumn{2}{c}{\includegraphics[width=0.192\textwidth]{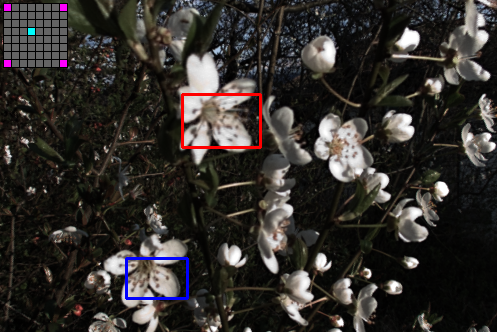}}
        & \multicolumn{2}{c}{\includegraphics[width=0.192\textwidth]{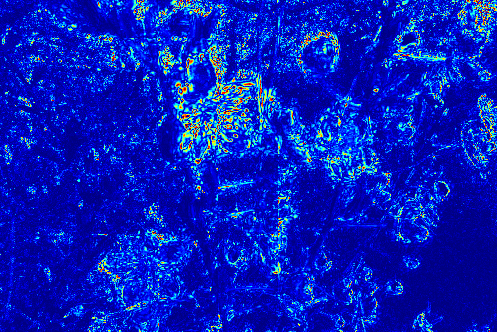}}
        & \multicolumn{2}{c}{\includegraphics[width=0.192\textwidth]{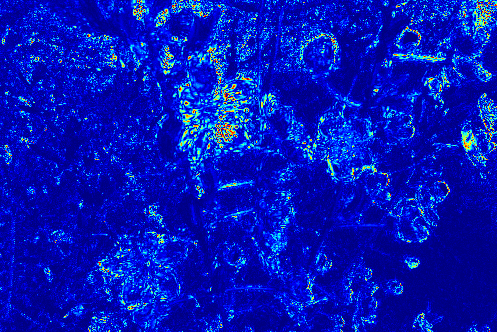}}
        & \multicolumn{2}{c}{\includegraphics[width=0.192\textwidth]{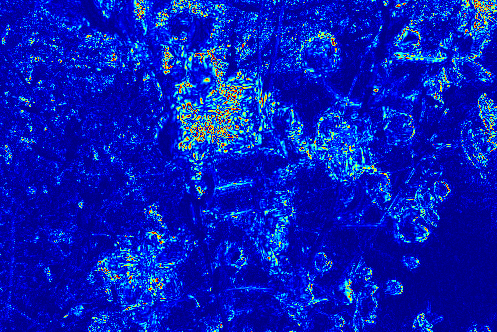}}
        & \multicolumn{2}{c}{\includegraphics[width=0.192\textwidth]{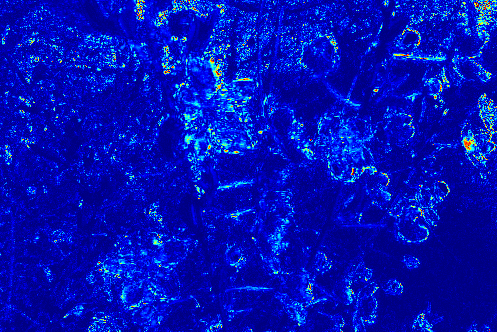}}
        \\

        \includegraphics[width=0.09\textwidth, height=0.052\textwidth, cfbox=red 1pt 0pt]{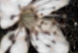}
        & \includegraphics[width=0.09\textwidth, height=0.052\textwidth, cfbox=blue 1pt 0pt]{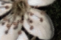}
        & \includegraphics[width=0.09\textwidth, height=0.052\textwidth, cfbox=red 1pt 0pt]{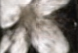}
        & \includegraphics[width=0.09\textwidth, height=0.052\textwidth, cfbox=blue 1pt 0pt]{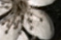}
        & \includegraphics[width=0.09\textwidth, height=0.052\textwidth, cfbox=red 1pt 0pt]{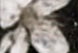}
        & \includegraphics[width=0.09\textwidth, height=0.052\textwidth, cfbox=blue 1pt 0pt]{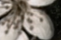}
        & \includegraphics[width=0.09\textwidth, height=0.052\textwidth, cfbox=red 1pt 0pt]{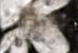}
        & \includegraphics[width=0.09\textwidth, height=0.052\textwidth, cfbox=blue 1pt 0pt]{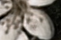}
        & \includegraphics[width=0.09\textwidth, height=0.052\textwidth, cfbox=red 1pt 0pt]{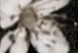}
        & \includegraphics[width=0.09\textwidth, height=0.052\textwidth, cfbox=blue 1pt 0pt]{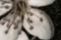}
        \\

        \multicolumn{2}{c}{\textit{Mirabelle\_Prune\_Tree}}
        & \multicolumn{2}{c}{31.05/0.9416}
        & \multicolumn{2}{c}{31.16/0.9509}
        & \multicolumn{2}{c}{30.63/0.9497}
        & \multicolumn{2}{c}{34.20/0.9743}
        \\
        \vspace{-0.15cm} \\

        \multicolumn{2}{c}{\includegraphics[width=0.192\textwidth]{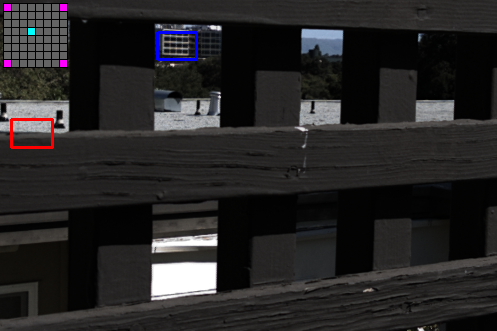}}
        & \multicolumn{2}{c}{\includegraphics[width=0.192\textwidth]{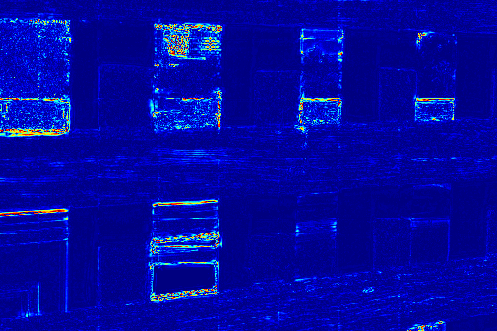}}
        & \multicolumn{2}{c}{\includegraphics[width=0.192\textwidth]{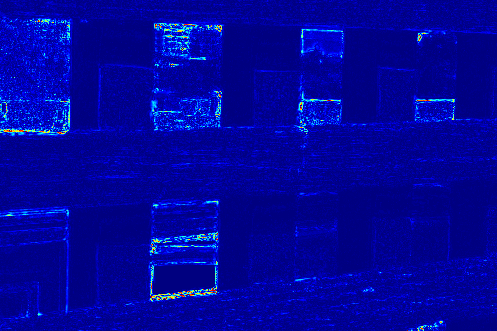}}
        & \multicolumn{2}{c}{\includegraphics[width=0.192\textwidth]{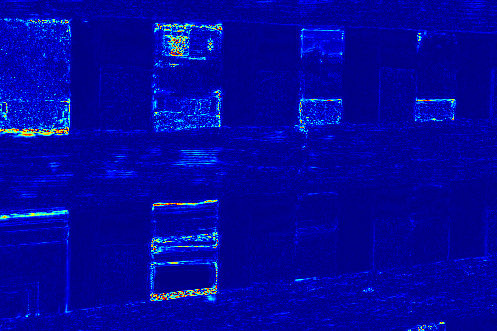}}
        & \multicolumn{2}{c}{\includegraphics[width=0.192\textwidth]{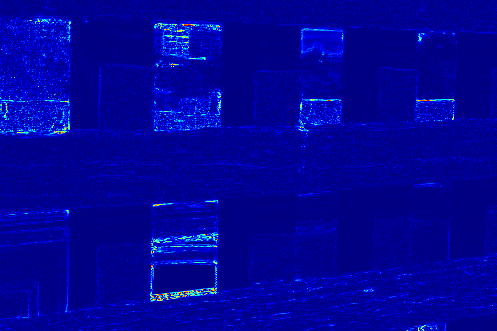}}
        \\

        \includegraphics[width=0.09\textwidth, height=0.052\textwidth, cfbox=red 1pt 0pt]{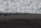}
        & \includegraphics[width=0.09\textwidth, height=0.052\textwidth, cfbox=blue 1pt 0pt]{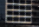}
        & \includegraphics[width=0.09\textwidth, height=0.052\textwidth, cfbox=red 1pt 0pt]{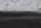}
        & \includegraphics[width=0.09\textwidth, height=0.052\textwidth, cfbox=blue 1pt 0pt]{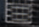}
        & \includegraphics[width=0.09\textwidth, height=0.052\textwidth, cfbox=red 1pt 0pt]{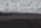}
        & \includegraphics[width=0.09\textwidth, height=0.052\textwidth, cfbox=blue 1pt 0pt]{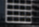}
        & \includegraphics[width=0.09\textwidth, height=0.052\textwidth, cfbox=red 1pt 0pt]{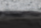}
        & \includegraphics[width=0.09\textwidth, height=0.052\textwidth, cfbox=blue 1pt 0pt]{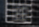}
        & \includegraphics[width=0.09\textwidth, height=0.052\textwidth, cfbox=red 1pt 0pt]{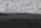}
        & \includegraphics[width=0.09\textwidth, height=0.052\textwidth, cfbox=blue 1pt 0pt]{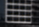}
        \\

        \multicolumn{2}{c}{\textit{Occlusions\_29}}
        & \multicolumn{2}{c}{34.58/0.9666}
        & \multicolumn{2}{c}{37.65/0.9800}
        & \multicolumn{2}{c}{36.33/0.9767}
        & \multicolumn{2}{c}{39.31/0.9872}
        \\
        \vspace{-0.15cm} \\

        \multicolumn{2}{c}{\includegraphics[width=0.192\textwidth]{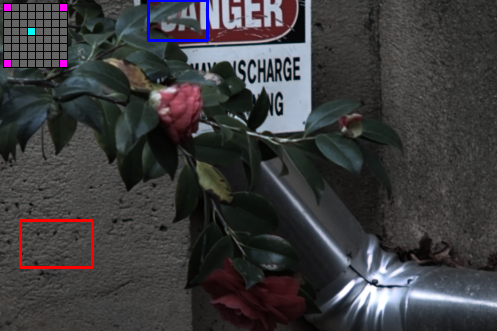}}
        & \multicolumn{2}{c}{\includegraphics[width=0.192\textwidth]{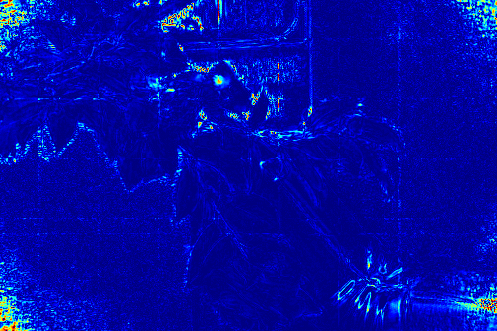}}
        & \multicolumn{2}{c}{\includegraphics[width=0.192\textwidth]{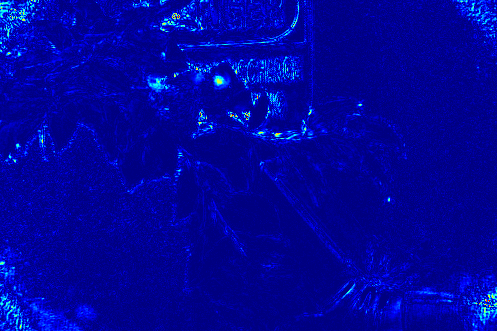}}
        & \multicolumn{2}{c}{\includegraphics[width=0.192\textwidth]{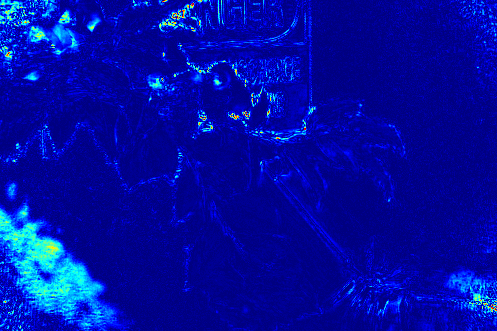}}
        & \multicolumn{2}{c}{\includegraphics[width=0.192\textwidth]{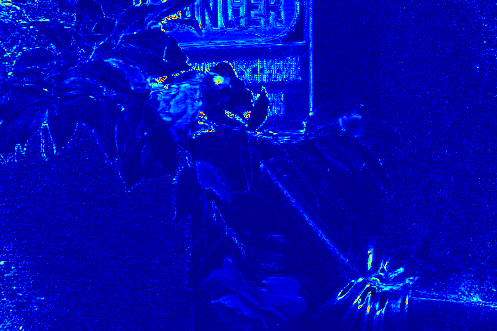}}
        \\

        \includegraphics[width=0.09\textwidth, height=0.052\textwidth, cfbox=red 1pt 0pt]{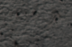}
        & \includegraphics[width=0.09\textwidth, height=0.052\textwidth, cfbox=blue 1pt 0pt]{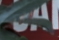}
        & \includegraphics[width=0.09\textwidth, height=0.052\textwidth, cfbox=red 1pt 0pt]{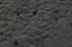}
        & \includegraphics[width=0.09\textwidth, height=0.052\textwidth, cfbox=blue 1pt 0pt]{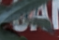}
        & \includegraphics[width=0.09\textwidth, height=0.052\textwidth, cfbox=red 1pt 0pt]{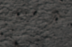}
        & \includegraphics[width=0.09\textwidth, height=0.052\textwidth, cfbox=blue 1pt 0pt]{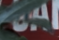}
        & \includegraphics[width=0.09\textwidth, height=0.052\textwidth, cfbox=red 1pt 0pt]{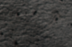}
        & \includegraphics[width=0.09\textwidth, height=0.052\textwidth, cfbox=blue 1pt 0pt]{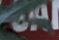}
        & \includegraphics[width=0.09\textwidth, height=0.052\textwidth, cfbox=red 1pt 0pt]{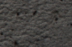}
        & \includegraphics[width=0.09\textwidth, height=0.052\textwidth, cfbox=blue 1pt 0pt]{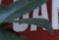}
        \\

        \multicolumn{2}{c}{\textit{Reflective\_12}}
        & \multicolumn{2}{c}{32.83/0.9494}
        & \multicolumn{2}{c}{34.98/0.9662}
        & \multicolumn{2}{c}{33.14/0.9635}
        & \multicolumn{2}{c}{36.10/0.9744}
        \\
        \vspace{-0.15cm} \\
        
    \end{tabular}
    
    \caption{Visualization of the reconstruction quality of the selected real-world scenes. The angular locations are indicated at the top-left corner in the ground truth. In each sample, the first row is pixel-wise error visualized as heat maps in the first row, the second row is zoom-in views of the selected regions in the red and blue boxes, and the third row is the sample title and PSNR/SSIM. }
    \label{fig:perfVis}
\end{figure*}

For a better understanding of the performance gains, selected visual results from the four test datasets are presented in Fig. \ref{fig:perfVis}. It is visible that our error maps are basically clearer than the other prior arts. Concretely, the reconstructed edges of the objects are more complete in \textit{IMG\_1528}, \textit{e.g.}, the lamp pole behind the leaves in the red bounding box and the sign above the vehicles in the blue bounding box. The lamp pole example poses a major challenge to LF reconstruction as it is occluded by the leaves in some SAIs. Such a phenomenon can also be observed at the fence border in the red bounding box of \textit{Occlusions\_29} and the English letter behind the leaves in the blue bounding box of \textit{Reflective\_12}. This can be attributed to the abundant spatio-angular representation in handling the appearance that is visible in limited SAIs. Similar improvement can also be observed in the reconstruction of complicated details, \textit{e.g.}, the flower cores in the blue and red boxes of \textit{Mirabelle\_Prune\_Tree}, and the windows in the blue bounding box of \textit{Occlusions\_29}, which are entirely blurred in the result of most methods. In \textit{Reflective\_12}, it is very obvious that the method of HDDRNet \cite{mengHDDRNet_TPAMI2019} and Yeung \textit{et al.} \cite{yeungSAS_ECCV2018} incorrectly reconstructs a large area of shadow at the bottom left corner while SADenseNet successfully reconstructs it with the long-term information supplied by the dense skip connections \cite{taiMemNet_ICCV2017} and enlarged spatial receptive fields of the correlation blocks while HDDRNet \cite{mengHDDRNet_TPAMI2019} and Yeung \textit{et al.} \cite{yeungSAS_ECCV2018} confine the correlation information and falsely renders the shadow across all SAIs.

On the other hand, to demonstrate the model's efficiency, we compare the memory cost in terms of the number of trainable parameters and the computation cost in terms of the testing speed, reported in Table \ref{tab:OverallPerformance}. The speed tests are operated in CPU-only mode. The results demonstrate that the method of Kalantari \textit{et al.} \cite{kalantari_SIGGRAPHASIA2016} and NoisyLFRecon \cite{ZhouNoisyLFRecon_SS2021} are plagued by their explicit disparity estimation and warping process. Also, their SAI synthesis functions must be performed separately for each SAI as the intermediate information cannot be shared ending up with a very low processing speed. Regarding the memory cost, HDDRNet \cite{mengHDDRNet_TPAMI2019} and NoisyLFRecon \cite{ZhouNoisyLFRecon_SS2021} employ a relatively huge model as the former hires 4D convolution operators and the latter adopts a modified ResNet-50 which has taken more than 19 million parameters. These two means aim at extracting an expressive feature representation but lead to a huge model instead of improvement in reconstruction performance. On the other hand, the end-to-end separable-convolution-based methods, \cite{yeungSAS_ECCV2018} and SADenseNet, reconstruct SAIs in one propagation at a substantially faster speed. Moreover, compared with \cite{yeungSAS_ECCV2018}, our SADenseNet requires only 3/4 of parameters and is nearly 3 times faster. The achievement can be attributed to the correlation blocks with efficient convolutions and spatio-angular dense skip connections that improve information flow.

\begin{figure*}[ht!]
    \centering
    \tabcolsep=0.02cm
    \renewcommand{\arraystretch}{0.5}
    \begin{tabular}{ccccccc}
        Ground truth &
        Yeung \textit{et al.} \cite{yeungSAS_ECCV2018} &
        SADenseNet (Ours) &
        {} &
        Ground truth &
        Yeung \textit{et al.} \cite{yeungSAS_ECCV2018} &
        SADenseNet (Ours) \\

        \includegraphics[width=0.16\textwidth]{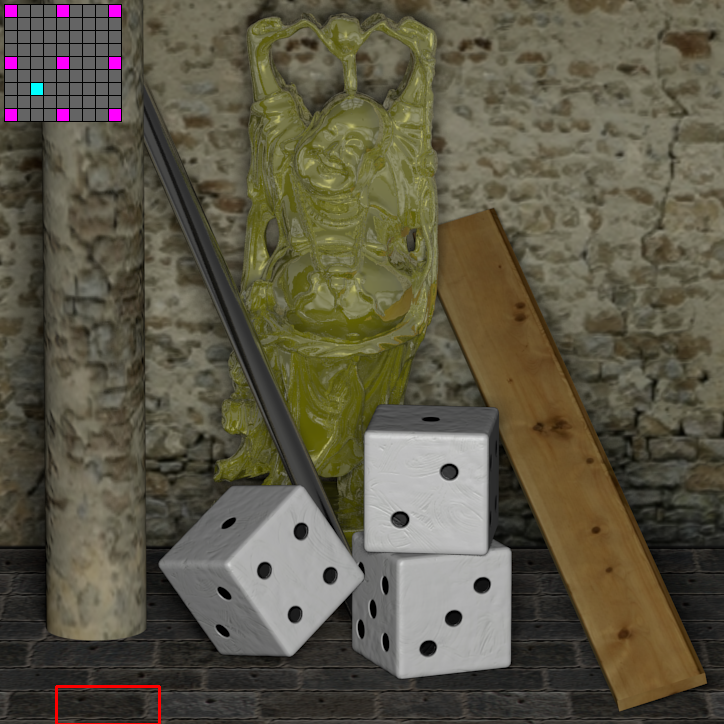} &
        \includegraphics[width=0.16\textwidth]{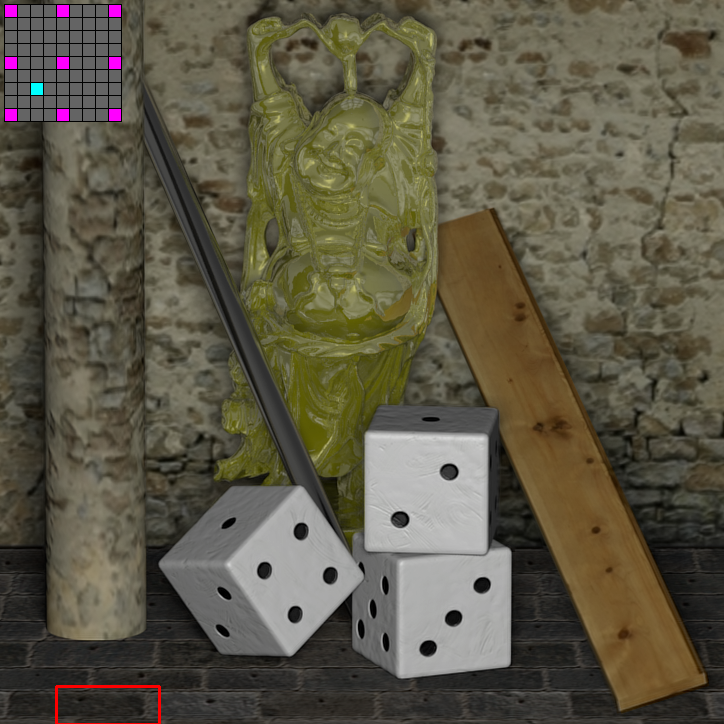} &
        \includegraphics[width=0.16\textwidth]{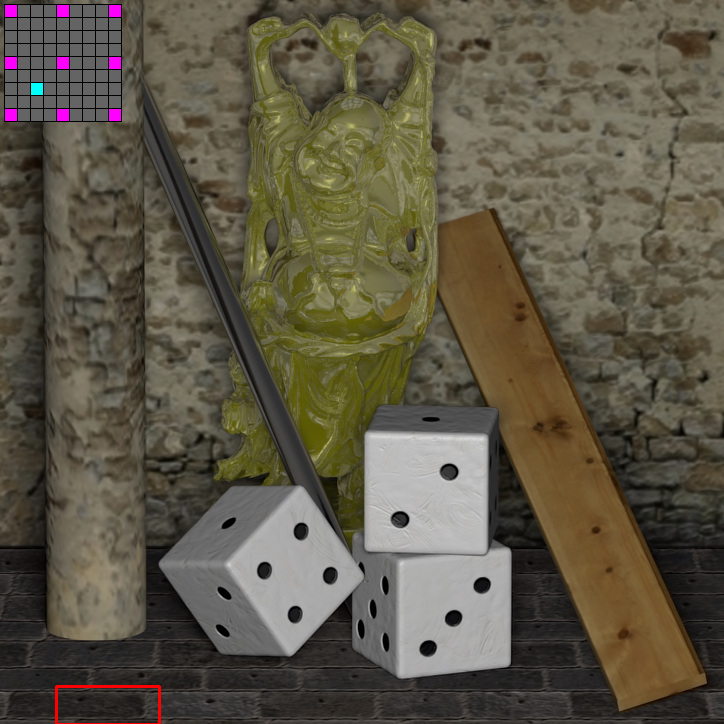} &
        \hspace{0.15cm} &
        \includegraphics[width=0.16\textwidth]{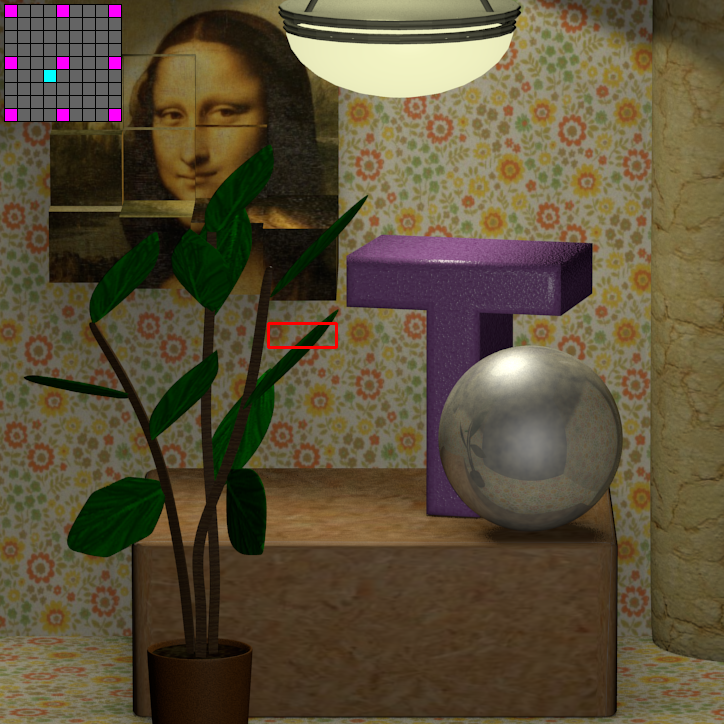} &
        \includegraphics[width=0.16\textwidth]{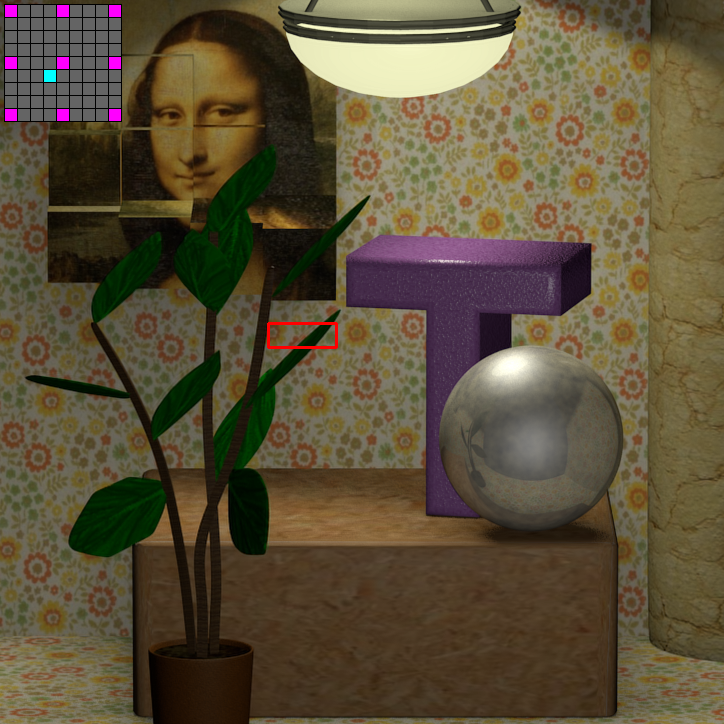} &
        \includegraphics[width=0.16\textwidth]{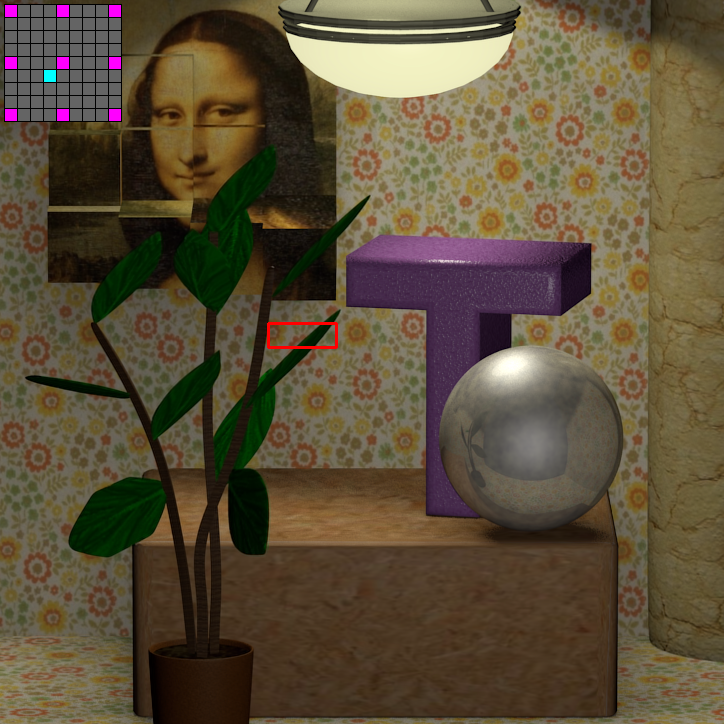} \\
        
        \includegraphics[width=0.155\textwidth, cfbox=red 1pt 0pt]{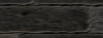} &
        \includegraphics[width=0.155\textwidth, cfbox=red 1pt 0pt]{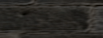} &
        \includegraphics[width=0.155\textwidth, cfbox=red 1pt 0pt]{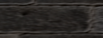} &
        {} &
        \includegraphics[width=0.155\textwidth, cfbox=red 1pt 0pt]{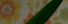} &
        \includegraphics[width=0.155\textwidth, cfbox=red 1pt 0pt]{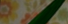} &
        \includegraphics[width=0.155\textwidth, cfbox=red 1pt 0pt]{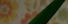} \\

        {} &
        \includegraphics[width=0.16\textwidth]{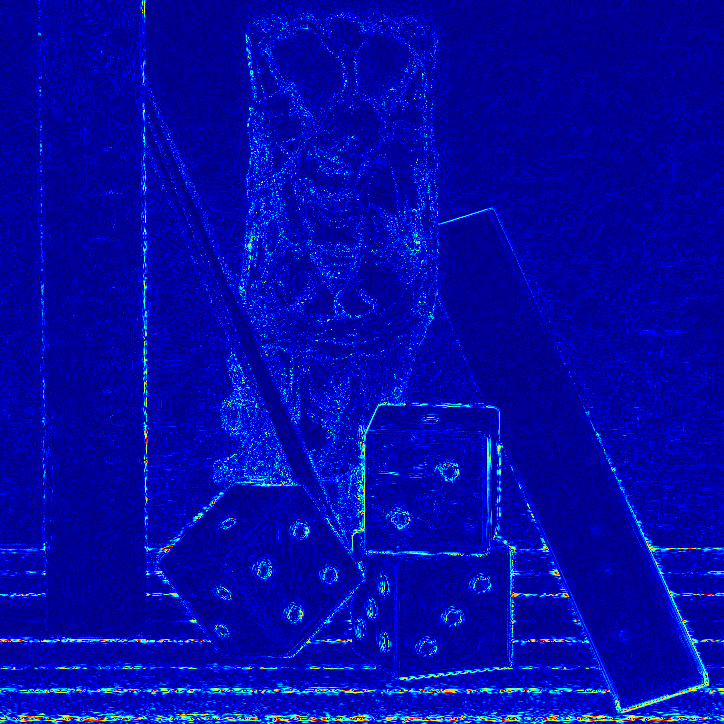} &
        \includegraphics[width=0.16\textwidth]{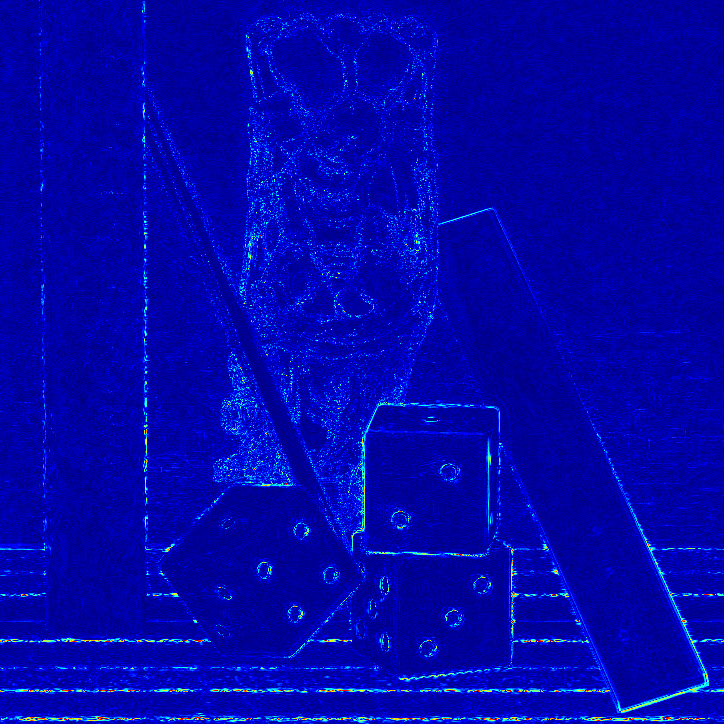} &
        {} &
        {} &
        \includegraphics[width=0.16\textwidth]{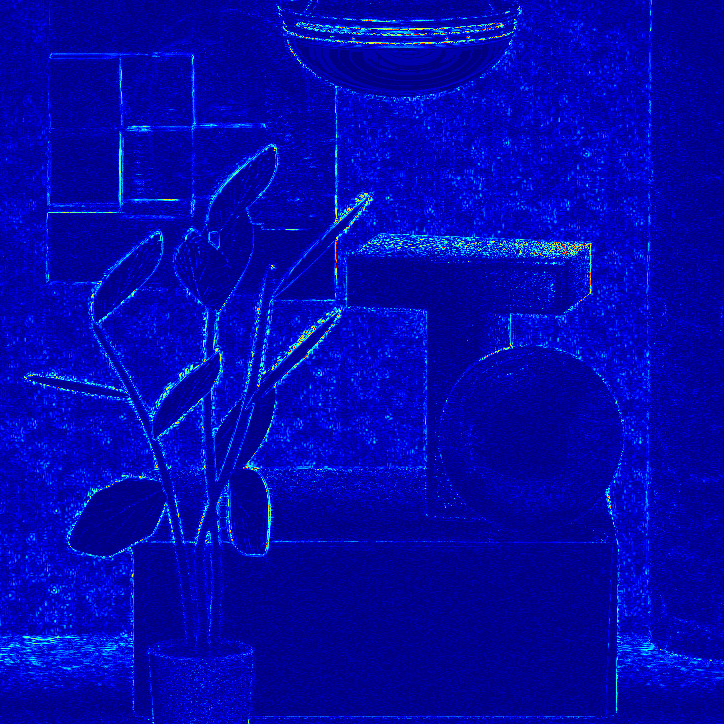} &
        \includegraphics[width=0.16\textwidth]{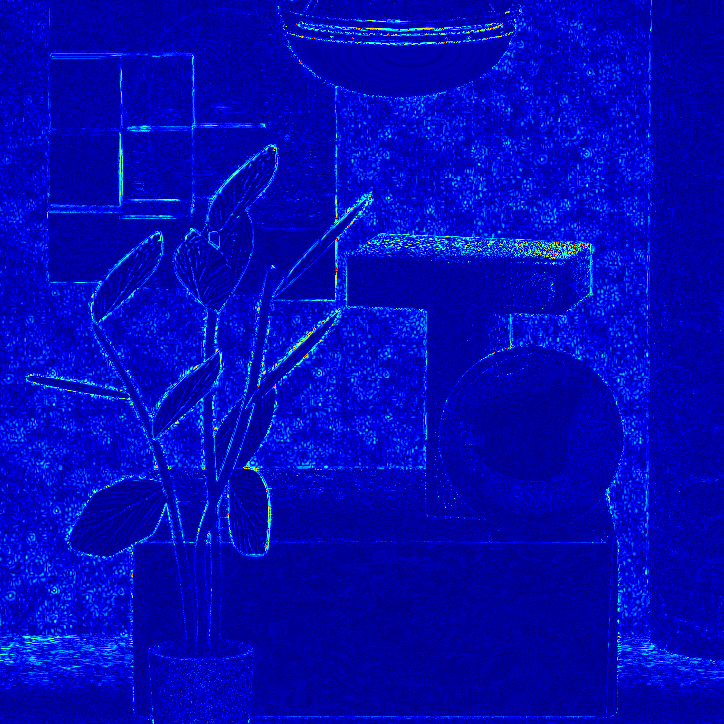} \\

        {} &
        43.77/0.9872 &
        45.82/0.9921&
        {} &
        {} &
        45.67/0.9920 &
        46.84/0.9932 \\

        \\

        \multicolumn{3}{c}{\textit{\small (a) Buddha}} &
        {} &
        \multicolumn{3}{c}{\textit{\small (b) Mona}}

    \end{tabular}
    \caption{Visualization of reconstruction quality of the selected scenes in the synthetic HCI dataset \cite{wannerHCI2013}. The two scenes, \textit{Buddha} and \textit{Mona}, are shown on the left and right respectively. In each sample, the three columns correspond to ground truth, the results of Yeung \textit{et al.} \cite{yeungSAS_ECCV2018} and SADenseNet. Zoom-in views of selected regions in the first row are displayed in the red boxes of the second row. Error maps are depicted as heat maps in the third row.}
    \label{fig:perfHCI}
\end{figure*}

\subsection{Comparison with State-of-the-art Methods on Synthetic Images}
The performance of SADenseNet is also evaluated on the synthetic HCI dataset \cite{wannerHCI2013}. We follow the practice of Wu \textit{et al.} \cite{wuEPICNN_TPAMI2018} and Yeung \textit{et al.} \cite{yeungSAS_ECCV2018} to calculate PSNR and SSIM on the Y channel only over the $(3 \times 3)$ to $(9 \times 9)$ reconstruction task. The results on \textit{Buddha} and \textit{Mona} are reported in Table \ref{tab:HCIPerformance}. SADenseNet has produced competitive results exceeding Yeung \textit{et al.} \cite{yeungSAS_ECCV2018} by 1.61 dB PSNR averagely. In regard to SSIM, SADenseNet has obtained the closest results to the ones reported by Wu \textit{et al.} in \cite{wuEPICNN_TPAMI2018} while outperforming its PSNR significantly likewise.

The reconstruction quality of the two testing images has been visualized in Fig. \ref{fig:perfHCI}. The most significant difference in \textit{Buddha} is located at the edges of the bricks on the floor where SADenseNet does not produce the blurring artifacts as \cite{yeungSAS_ECCV2018}. Similar improvement has been observed in \textit{Mona} where artifacts are produced around the leaf border in \cite{yeungSAS_ECCV2018} but not seen in SADenseNet.

\subsection{Ablation Studies}
To further study the nature of SADenseNet, a series of ablation studies are conducted in this section. We keep the previous configuration in Section \ref{section: ExpRealWorld} to conduct the studies on real-world datasets.

\begin{table}[t]
    \centering
    \caption{Comparison of performance with different $n_{\mathcal{S}}$. The baseline variant is underlined.}
\label{table:ns}
\begin{tabu}{|c|[2pt]c|[2pt]c|c|}
\hline
$n_{\mathcal{S}}$   & 30 Scenes                                     & \# Parameters         & Speed/s               \\ \hline\hline
6                   & 40.34/0.9839                                  & 1,466,108               & 16.25                 \\ \hline
\uline{5}       & \uline{40.31/0.9836}                      & \uline{1,134,140}   & \uline{12.82}     \\ \hline
4                   & 40.15/0.9831                                  & 857,468                & 9.96                  \\ \hline
3                   & 40.01/0.9824                                  & 636,092                & 7.30                  \\ \hline
2                   & 39.80/0.9815                                  & 470,012                & 5.72                  \\ \hline
1                   & 38.28/0.9787                                  & 359,228                & 4.18                  \\ \hline
\end{tabu}
    \bigskip
    \caption{Comparison of performance with different $n_{\mathcal{A}}$. The baseline variant is underlined.}
\label{table:na}
\begin{tabu}{|c|[2pt]c|[2pt]c|c|}
\hline
$n_{\mathcal{A}}$   & 30 Scenes                                     & \# Parameters         & Speed/s               \\ \hline\hline
\uline{1}       & \uline{40.31/0.9836}                      & \uline{1,134,140}   & \uline{12.82}     \\ \hline
2                   & 40.27/0.9834                                  & 1,189,628               & 13.48                 \\ \hline
3                   & 40.19/0.9834                                  & 1,245,116               & 14.26                 \\ \hline
\end{tabu}
    \bigskip
    \caption{Comparison of performance with different $n_{CB}$. The baseline variant is underlined.}
\label{table:nc}
\begin{tabu}{|c|[2pt]c|[2pt]c|c|}
\hline
$n_{CB}$        & 30 Scenes                                     & \# Parameters         & Speed/s               \\ \hline\hline
7               & 40.31/0.9835                                  & 1,346,588               & 14.98                 \\ \hline
\uline{6}   & \uline{40.31/0.9836}                      & \uline{1,134,140}   & \uline{12.82}     \\ \hline
5               & 40.21/0.9834                                  & 930,908                & 10.52                 \\ \hline
4               & 40.03/0.9824                                  & 736,892                & 8.36                  \\ \hline
3               & 39.91/0.9819                                  & 552,092                & 6.48                  \\ \hline
2               & 39.61/0.9804                                  & 376,508                & 4.30                  \\ \hline
1               & 38.35/0.9735                                  & 210,240                & 2.41                  \\ \hline
\end{tabu}
\end{table}
\begin{figure}[t]
    \includegraphics[width=0.48\textwidth]{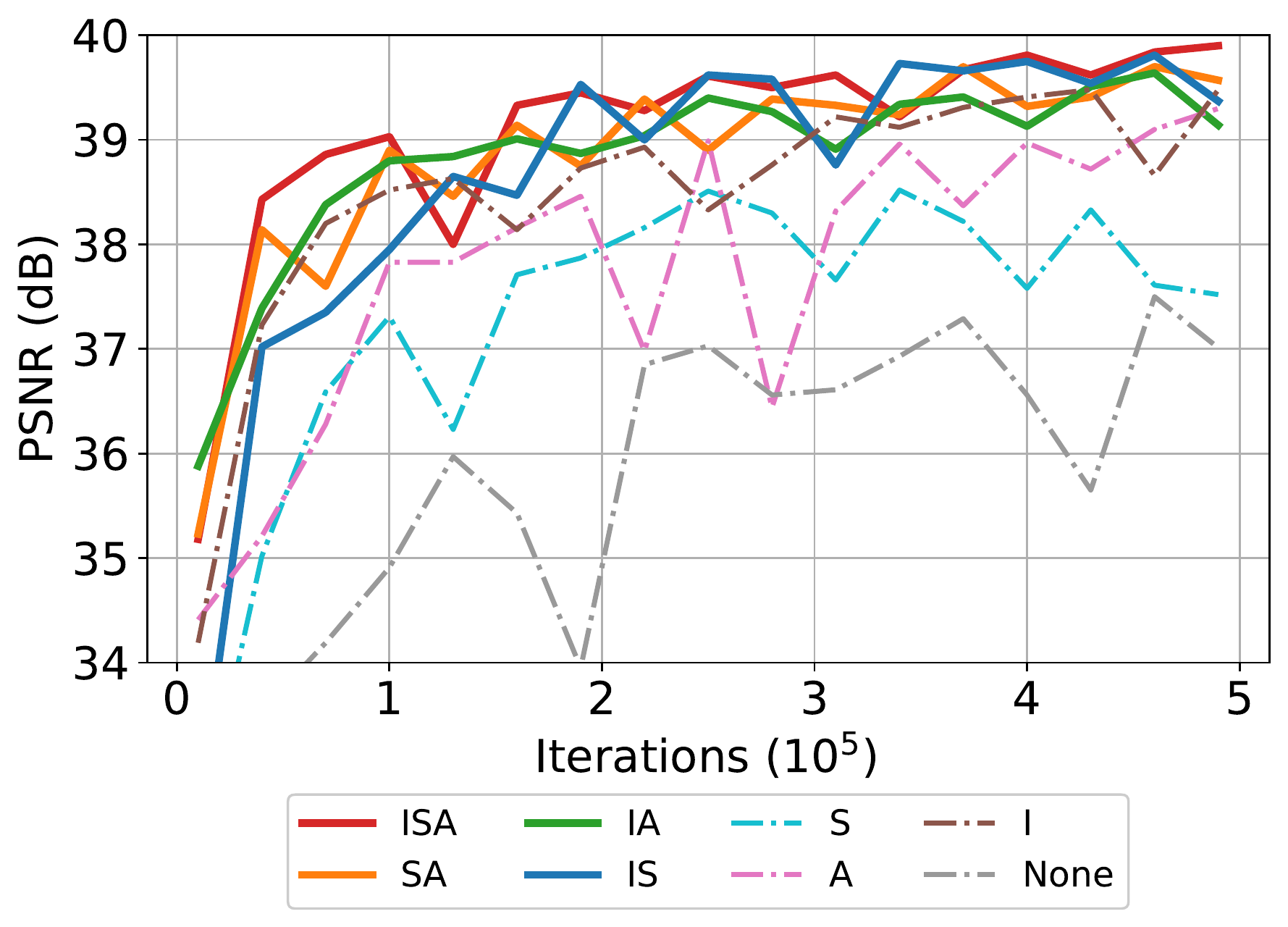}
    \caption{The training process of SADenseNet variants. The first $5 \times 10^5$ iterations are plotted. The variants with multiple connections are indicated by solid lines while the ones with a single connection or none are indicated by dashed lines. For visual effect, some curves are partially truncated.}
    \label{fig:variantsHistory}
\end{figure}

\subsubsection{Spatio-angular Dense Skip Connections}

\begin{figure*}[ht!]
    \centering
    \begin{tabular}{@{}cc@{}}
        \includegraphics[totalheight=6cm]{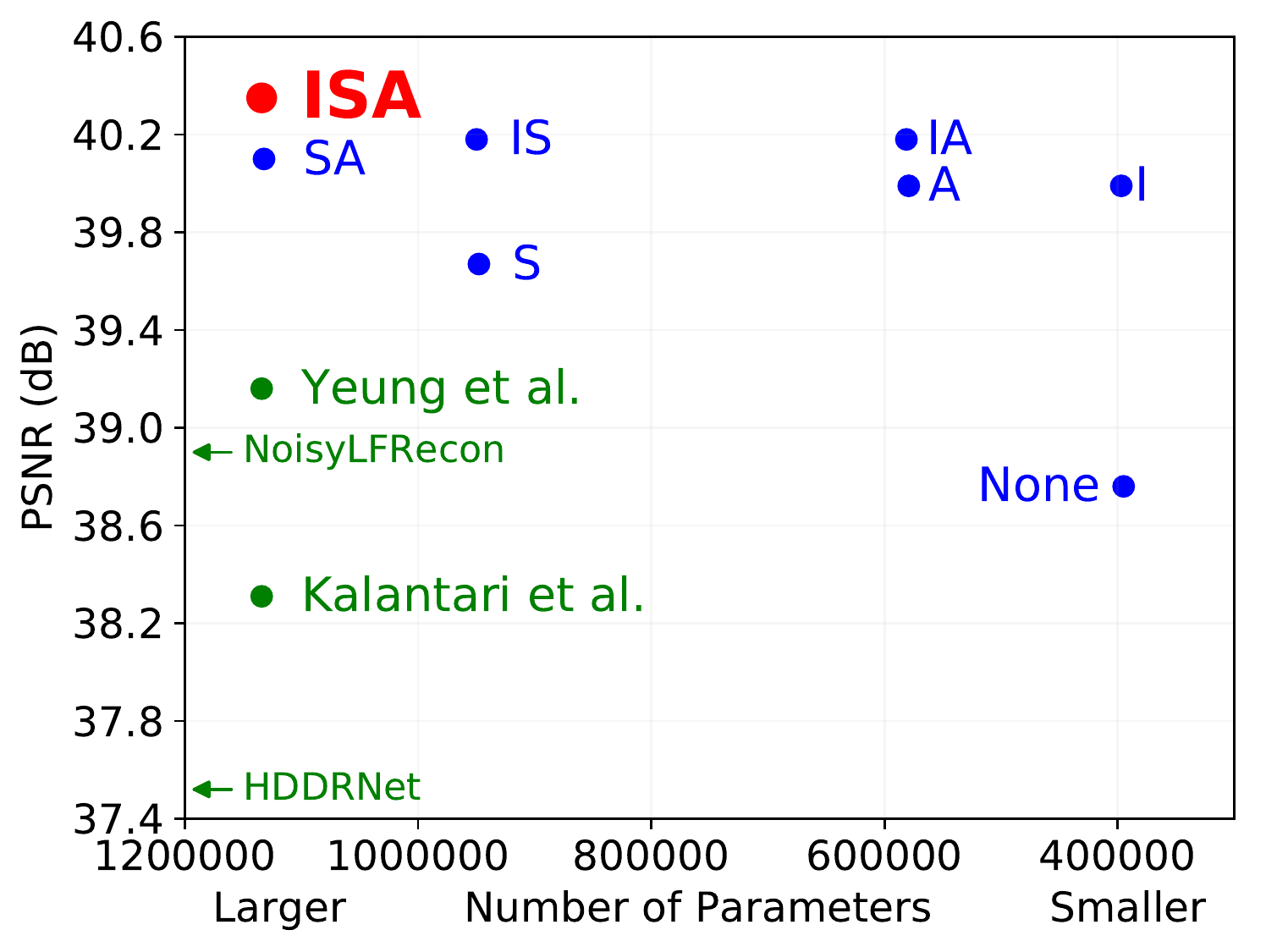} &
        \includegraphics[totalheight=6cm]{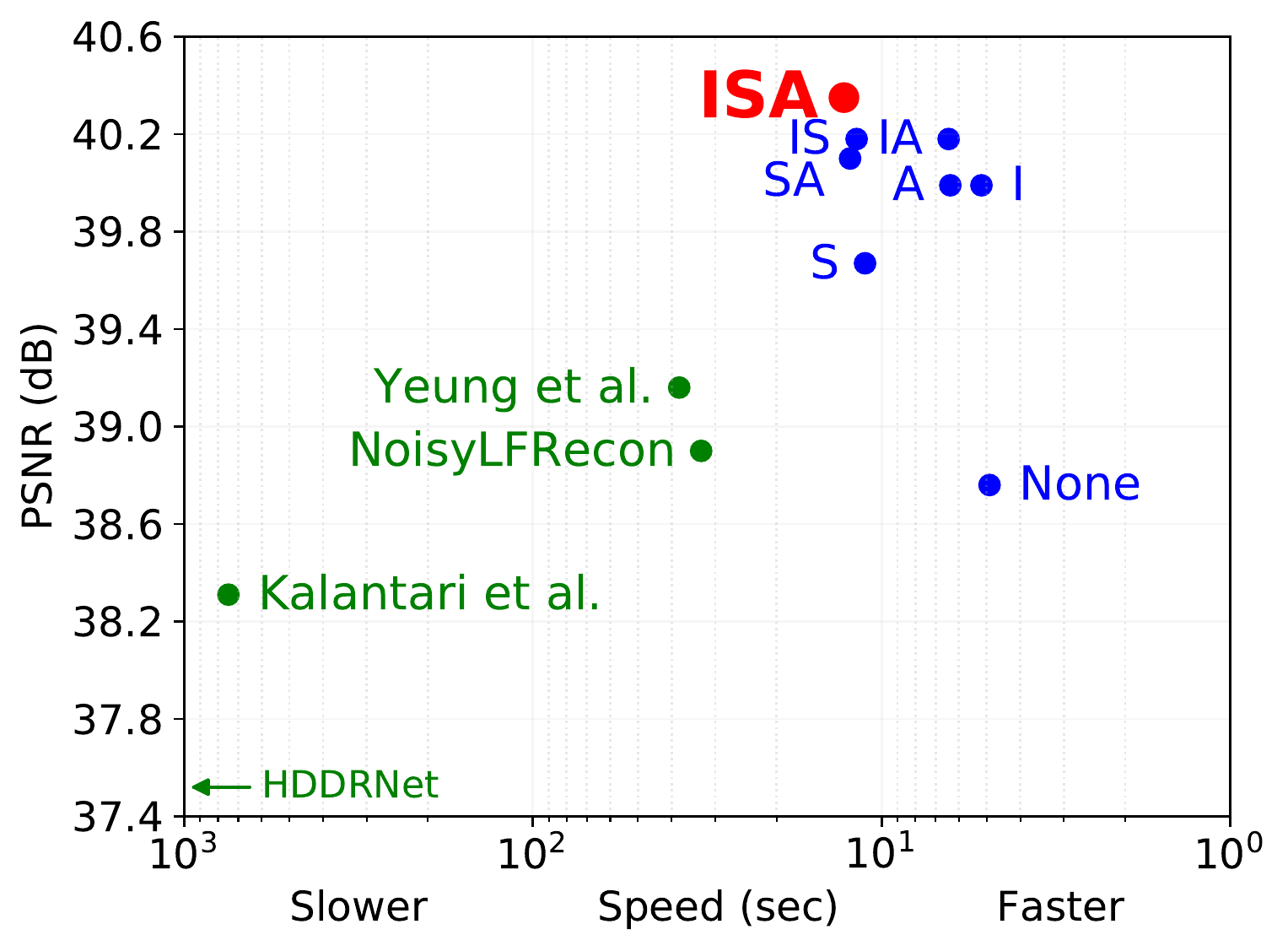} \\
        \small (a) PSNR versus numbers of parameters. &
        \small (b) PSNR versus speed.
    \end{tabular}
    \caption{The plots of the trade-off between reconstruction quality and costs of the SADenseNet variants and rivals. PSNR evaluated on 30 Scenes \cite{kalantari_SIGGRAPHASIA2016} is depicted as the metric of performance. The variants are in blue while the baseline SADenseNet\_ISA is in red. The compared methods are in green. For simplicity, the variants' prefix is omitted, and some items are indicated with arrows instead of dots as they are out of the scope. Generally, items in the top-right corners can handle the trade-off better with better performance and at smaller costs simultaneously.}
    \label{fig:perfVariants}
\end{figure*}

In Section \ref{section:Spatio-Angular Dense Skip Connection}, we have introduced three kinds of dense skip connections to enhance the information flow. To discover the contribution of these connections, we create several variants with or without them. For simplicity, spatial, angular and image connections are denoted as \textit{S}, \textit{A} and \textit{I} correspondingly. 8 variants, each processed a unique combination of connections, are reported in Table \ref{table:Variants}. SADenseNet\_ISA is effectively the full version of SADenseNet. $n_{CB}$ and $n_{\mathcal{S}}$ remain 6 and 5 correspondingly for all the variants. In Table \ref{table:Variants}, it is obvious that reconstruction quality deteriorates if any connection is removed. For a better comparison of the connections' contribution, we visualize these variants' performance against model sizes and speed in Fig. \ref{fig:perfVariants}. We set the full version, SADenseNet\_ISA, as the baseline in this study, and the results of the prior arts, Kalantari \textit{et al.} \cite{kalantari_SIGGRAPHASIA2016} and Yeung \textit{et al.} \cite{yeungSAS_ECCV2018}, are also plotted.

\begin{table*}[ht]
    \centering
    \caption{Performance comparison of SADenseNet variants on the real-world datasets. The table is divided into three parts: 1. The 2nd to 4th columns are the properties of the variants. 2. The 5th to 8th columns are PSNR and SSIM evaluated on four datasets. 3. The last two columns display the number of parameters and running speed. The baseline variant is underlined.}
    \label{table:Variants}
    \tabcolsep=0.15cm
    \begin{tabu}{|l|[2pt]c|c|c|[2pt]c|c|c|c|[2pt]c|c|}
        \hline
        Name                        &  Spatial  &   Angular &   Image   &   30 Scenes (30)              & EPFL (118)                & Occlusions (43)               & Reflective (31)               & \# Parameters             & Speed/s               \\ \hline\hline
        SADenseNet\_None            &           &           &           &   38.76/0.9776                & 39.25/0.9642              & 32.63/0.9123                  & 36.25/0.9453                  & 394,844                    & 4.9                   \\ \hline
        SADenseNet\_I               &           &           &   \cmark  &   39.85/0.9818                & 40.17/0.9683              & 33.30/0.9221                  & 36.90/0.9494                  & 396,860                    & 5.17                  \\ \hline
        SADenseNet\_S               &   \cmark  &           &           &   39.67/0.9815                & 39.89/0.9678              & 33.19/0.9199                  & 36.79/0.9492                  & 947,804                    & 11.15                 \\ \hline
        SADenseNet\_A               &           &   \cmark  &           &   39.99/0.9822                & 40.21/0.9686              & 33.40/0.9207                  & 36.99/0.9495                  & 579,164                    & 6.35                  \\ \hline
        SADenseNet\_SA              &   \cmark  &   \cmark  &           &   40.10/0.9828                & 40.30/0.9692              & 33.49/0.9241                  & 37.08/0.9496                  & 1,132,124                   & 12.31                 \\ \hline
        SADenseNet\_IA              &           &   \cmark  &   \cmark  &   40.18/0.9831                & 40.42/0.9701              & 33.67/0.9274                  & 37.06/0.9507                  & 581,180                    & 6.42                  \\ \hline
        SADenseNet\_IS              &   \cmark  &           &   \cmark  &   40.14/0.9832                & 40.32/0.9699              & 33.53/0.9238                  & 37.05/0.9500                  & 949,820                    & 11.79                 \\ \hline
        \uline{SADenseNet\_ISA} &   \cmark  &   \cmark  &   \cmark  &   \uline{40.31/0.9836}    & \uline{40.54/0.9706}  & \uline{33.76/0.9267}      & \uline{37.15/0.9521}      & \uline{1,134,140}       & \uline{12.82}     \\ \hline
    \end{tabu}

\end{table*}

In terms of reconstruction quality, while all types of connections outperform SADenseNet\_None significantly, the biggest performance gain is achieved by the angular connections as SADenseNet\_A surpasses SADenseNet\_S and SADenseNet\_I by 0.32 dB and 0.14 dB in 30 Scenes.

It is observed that the three kinds of dense skip connections are complementary as combining any two or three of them leads to further performance gains. Besides, all variants have outperformed the prior arts by at least 0.40 dB PSNR except SADenseNet\_None which is without any skip connections inducing drastic decline, proving the importance of the three kinds of skip connections.

In Fig. \ref{fig:perfVariants} (a) and (b), in terms of the x-axis, it is observed that all the SADenseNet variants have fewer parameters and higher speeds. While the prior arts take 32.92 seconds in \cite{ZhouNoisyLFRecon_SS2021}, 38.05 seconds in \cite{yeungSAS_ECCV2018} and even more than 700 seconds in \cite{kalantari_SIGGRAPHASIA2016, mengHDDRNet_TPAMI2019} to reconstruct a full LF image, SADenseNet variants take less than half of the time. It is worth noticing that while SADenseNet\_ISA has achieved the best reconstruction quality with a large model size similar to Yeung \textit{et al.} \cite{yeungSAS_ECCV2018}, the sub-optimal variant SADenseNet\_IA is capable of achieving approximate performance, 40.18 dB PSNR, 0.13 dB lower than SADenseNet\_ISA, but it uses only 0.58 million parameters, which is about half of SADenseNet\_ISA, manifesting to be an economic option under a tight memory constraint. The speed of SADenseNet\_IA is about 2$\times$ faster than the variants with spatial connections as it gets rid of the relatively costly spatial connections. Meanwhile, SADenseNet\_I is also a considerable choice to pursue a lightweight model under a tighter condition which takes a similar number of parameters and running speed with SADenseNet\_None but achieves 39.85 dB PSNR, 1.09 dB higher.

Last but not least, the training process is plotted in Fig. \ref{fig:variantsHistory} for demonstrating how these connections improve the information flow and benefit the convergence of the network during training. For simplicity, we only plot the first $5 \times 10^5$ iterations with PSNR. The curves appear more stable when multiple connections are used, \textit{i.e.} SADenseNet\_ISA, SADenseNet\_SA, SADenseNet\_IA, and SADenseNet\_IS, while the ones with either spatial connections, angular connections, or none have encountered sudden drops in PSNR. We deduce that the improvement of training stability is derived from the combination of spatial and angular dense connected information flow. However, SADenseNet\_I is the only exception having a stable curve, suggesting that this type of connection is complementary to the extracted spatio-angular features while training. This impact is also verified by comparing SADenseNet\_IA and SADenseNet\_IS with SADenseNet\_A and SADenseNet\_S. It is also witnessed that SADenseNet\_ISA, trained with $1.5 \times 10^5$ iterations in around 14 hours, has already outperformed the best result of Yeung \textit{et al.} \cite{yeungSAS_ECCV2018} which requires 10 days of training on a similar GPU. 

With the analysis from the test scenario and the training process, we can draw a conclusion that the proposed three kinds of dense skip connections can enhance the information flow of SADenseNet to facilitate a quick and stable convergence when training and improve the reconstruction performance while the model remains efficient in computation and memory usage.

\begin{figure*}[ht!]
    \centering
    \tabcolsep=0.04cm
    \begin{tabular}{ccccc}
        & Central SAI & Depth from ground truth & Depth from Yeung \textit{et al.} \cite{yeungSAS_ECCV2018} & Depth from SADenseNet \\

        \rotatebox{90}{\hspace{33pt}\textit{Cars}} &
        \includegraphics[width=0.24\textwidth]{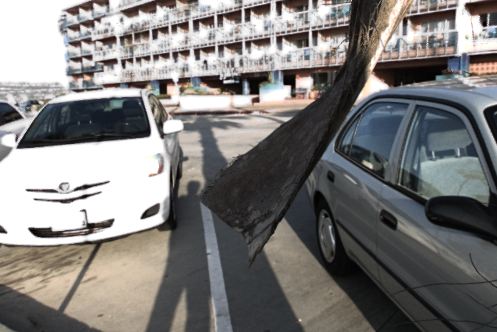} &
        \includegraphics[width=0.24\textwidth]{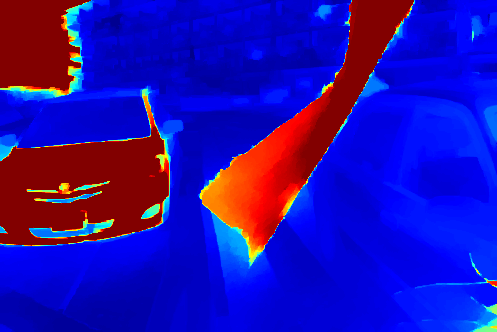} &
        \includegraphics[width=0.24\textwidth]{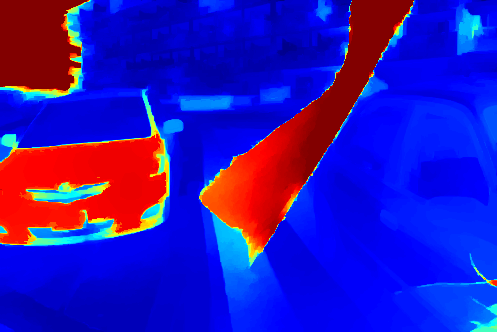} &
        \includegraphics[width=0.24\textwidth]{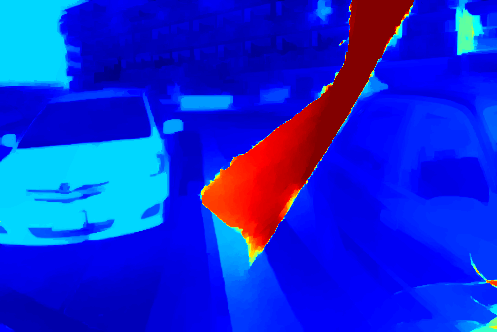} \\

        \rotatebox{90}{\hspace{20pt}\textit{IMG\_1411}} &
        \includegraphics[width=0.24\textwidth]{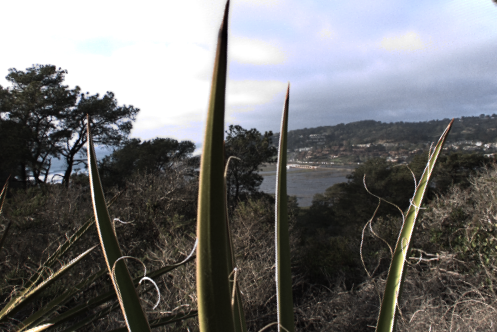} &
        \includegraphics[width=0.24\textwidth]{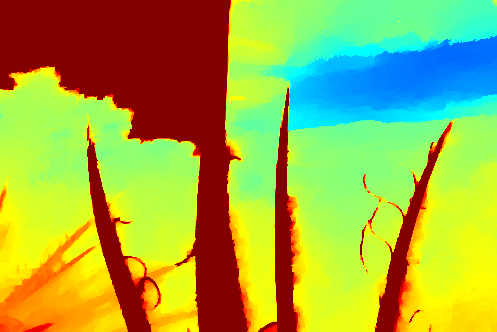} &
        \includegraphics[width=0.24\textwidth]{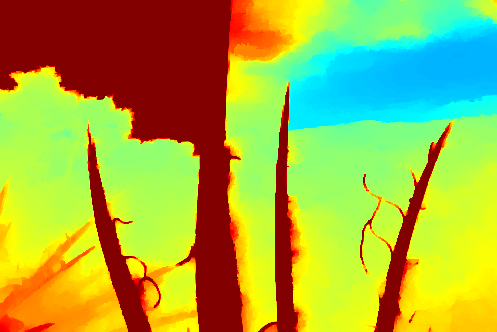} &
        \includegraphics[width=0.24\textwidth]{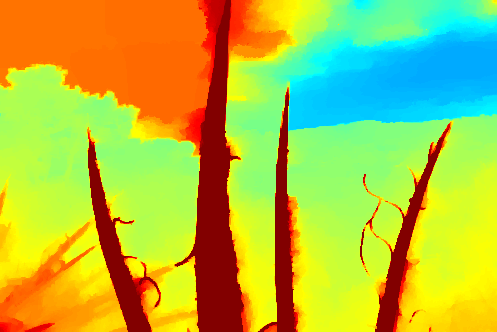} \\

        \rotatebox{90}{\hspace{10pt}\footnotesize\textit{Chain-link\_Fence\_2}} &
        \includegraphics[width=0.24\textwidth]{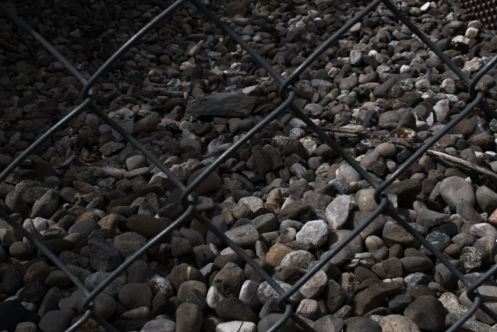} &
        \includegraphics[width=0.24\textwidth]{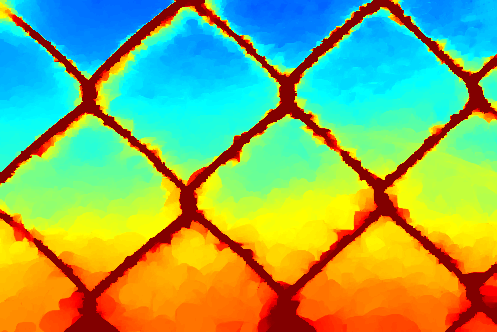} &
        \includegraphics[width=0.24\textwidth]{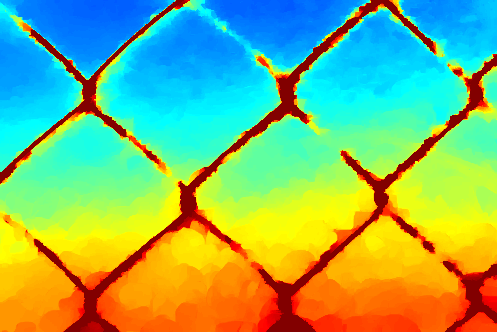} &
        \includegraphics[width=0.24\textwidth]{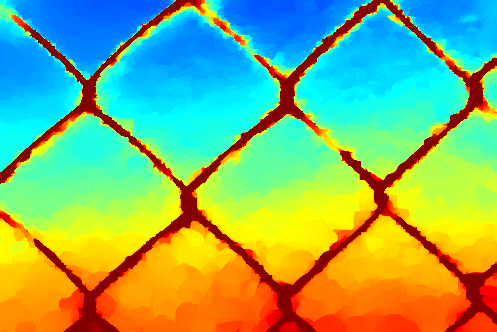} \\
    \end{tabular}
    \caption{The estimated depth of selected examples. The first column is the raw image information followed by three columns with depth maps estimated by the method of Chen \textit{et al.} \cite{chen2018accurate} on the full LF of the ground truth, the reconstructed LF of Yeung \textit{et al.} \cite{yeungSAS_ECCV2018} and SADenseNet.}
    \label{fig:depth}
\end{figure*}

\subsubsection{Domain Asymmetry} \label{section:Ablation Study Asymmetry}
As described in Section \ref{section:Introduction}, we hypothesize the domain asymmetry in the spatial and angular domains. In this section, we validate this hypothesis by evaluating the performance with different $n_\mathcal{S}$ on 30 Scenes \cite{kalantari_SIGGRAPHASIA2016} dataset. The number of correlation blocks $n_{CB}$ is fixed to 6 and we set $n_\mathcal{S}=5$ as the baseline. As shown in Table \ref{table:ns}, the performance reduces gradually as $n_{\mathcal{S}}$ decreases. It is worth noticing that when $n_{\mathcal{S}}$ decreases to 1, a correlation block virtually degrades to a SAS convolution as in \cite{yeungSAS_ECCV2018}, which achieves the worst results of 38.28 dB, declining by 2.03 dB, suggesting it hampers the performance by treating the two domains symmetrically. Moreover, when adding more spatial convolutions to the baseline, the performance gain is marginal. When $n_{\mathcal{S}}=6$, PSNR increases by merely 0.03 dB but 29\% more parameters and about 22\% lower speed are incurred. This suggests that the influence of deeper spatial features has saturated.

On the other hand, we modify the correlation blocks to append more angular convolutions. The number of angular convolutions in each block is denoted as $n_{\mathcal{A}}$, and the other variables $n_{CB}$ and $n_{\mathcal{S}}$ are fixed to 6 and 5. In each block, only the last angular convolution is densely connected with other blocks. Such a modification seems likely to obtain deeper angular features, however, the results in Table \ref{table:na} disagree with the speculation. Harmful effects are recorded that when $n_{\mathcal{A}}=2$ and $3$, the PSNR declines by 0.04 dB and 0.12 dB compared with the baseline, let alone the incurred computation and memory cost, proving that employing more angular convolutions is in vain.

Given the results on $n_{\mathcal{S}}$ and $n_{\mathcal{A}}$, we can conclude that domain asymmetry exists in LF data, and the correlation blocks are proved effective to extract spatio-angular features with asymmetrical operations.

\subsubsection{Correlation Blocks}
To verify the effectiveness of stacking correlation blocks, we evaluate the performance by varying the number of blocks $n_{CB}$. As demonstrated in Table \ref{table:nc}, the performance is rising as $n_{CB}$ increases and reaches the highest point at $n_{CB}=6$, which is used as the baseline. When $n_{CB}$ increases to 7, the performance slightly drops, which suggests the performance gain brought by a deeper architecture has been saturated, and adding more correlation blocks will lead to deterioration.

Remarkably, the $n_{CB}=2$ variant has outperformed Yeung \textit{et al.} \cite{yeungSAS_ECCV2018} by 0.45 dB PSNR with only 33\% parameters and $8.8\times$ faster speed, another convincing proof of the architecture's effectiveness and another option to obtain a lightweight but powerful model by employing fewer correlation blocks.

\subsection{Depth Estimation}
Depth estimation is one of the important LF measurement applications. As SAIs are reconstructed, we further investigate the depth maps estimated from the densely sampled LF to demonstrate SADenseNet's reconstruction quality from another perspective. Fig. \ref{fig:depth} gives some selected images' depth maps generated from the ground truth, the densely-sampled LF reconstructed by Yeung \textit{et al.} \cite{yeungSAS_ECCV2018} and SADenseNet using the estimation method of \cite{chen2018accurate}. It can be observed that SADenseNet has produced visibly better depth maps than Yeung \textit{et al.} \cite{yeungSAS_ECCV2018} with high-quality object edges in \textit{Chain-link\_Fence\_2}. It is interesting that in \textit{Cars} and \textit{IMG\_1411}, depth generated on SADenseNet's reconstructed SAIs is more satisfactory than ground truth. In \textit{Cars}, the car on the left should be located between the metal plate in the foreground and the building in the background. Both ground truth and Yeung \textit{et al.} \cite{yeungSAS_ECCV2018} misclassify it as having a similar depth to the metal plate. The misclassification mainly happens in textureless areas such as the sky in the top-left corner of \textit{Cars} and to the left of the plant in \textit{IMG\_1411}. However, depth maps estimated with SADenseNet do not suffer from this problem. 

The depth predictions in Fig. \ref{fig:depth} suggests that the region of the textureless areas and the foreground objects become more distinguishable to the depth estimation algorithm \cite{chen2018accurate} after being pre-processed by our proposed method. One plausible explanation for such improvement is that the prior learned by our proposed method is able to refine the input LF during reconstruction before feeding into the depth estimation algorithm. In conclusion, SADenseNet not only produces high reconstruction quality but also supplies complementary information to improve the accuracy of depth estimation and other measurement applications. 

\begin{figure*}[ht!]
    \centering
    \tabcolsep=0.02cm
    \renewcommand{\arraystretch}{0.5}
    \begin{tabular}{ccccccccc}
        \small Ground truth &
        \small NoisyLFRecon &
        \small Yeung \textit{et al.} &
        \small SADenseNet &
        {} &
        \small Ground truth &
        \small NoisyLFRecon &
        \small Yeung \textit{et al.} &
        \small SADenseNet \\

        {} &
        \small \cite{ZhouNoisyLFRecon_SS2021} &
        \small \cite{yeungSAS_ECCV2018} &
        \small (Ours) &
        {} &
        {} &
        \small \cite{ZhouNoisyLFRecon_SS2021} &
        \small \cite{yeungSAS_ECCV2018} &
        \small (Ours) \\

        \includegraphics[width=0.12\textwidth]{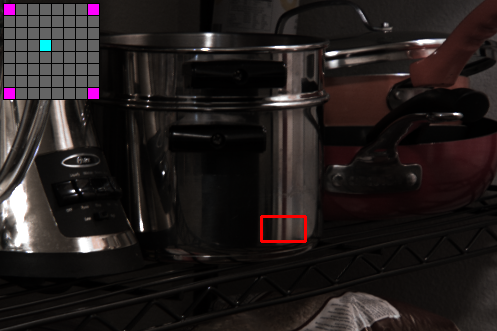} &
        \includegraphics[width=0.12\textwidth]{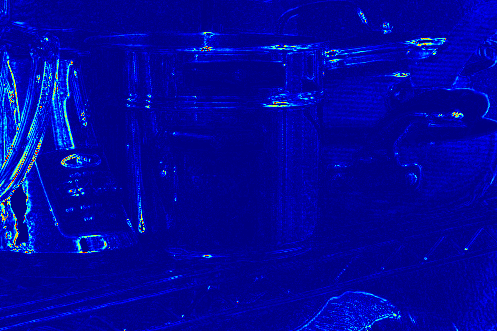} &
        \includegraphics[width=0.12\textwidth]{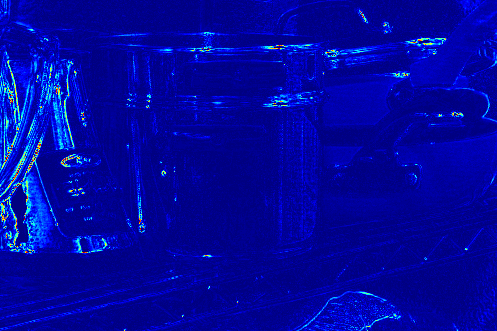} &
        \includegraphics[width=0.12\textwidth]{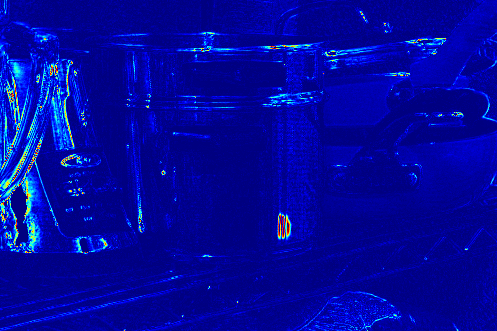} &
        \hspace{0.15cm} &
        \includegraphics[width=0.12\textwidth]{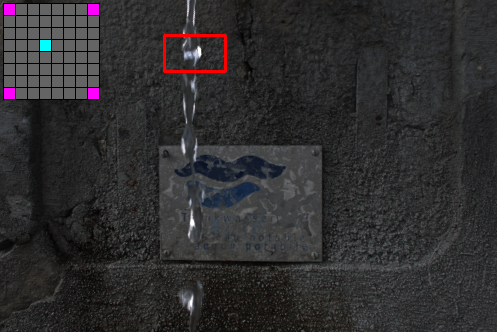} &
        \includegraphics[width=0.12\textwidth]{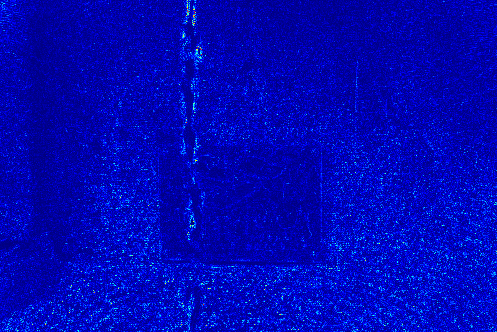} &
        \includegraphics[width=0.12\textwidth]{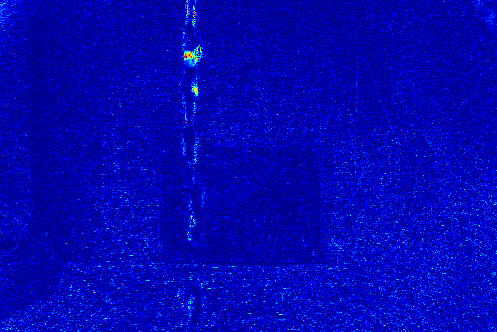} &
        \includegraphics[width=0.12\textwidth]{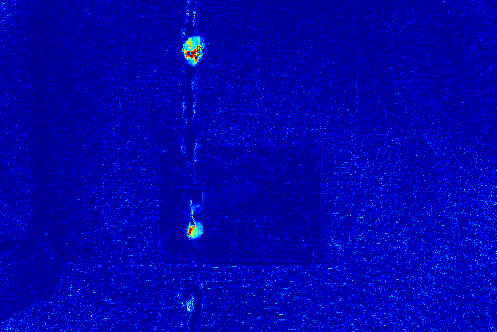} \\

        \includegraphics[width=0.115\textwidth, height=0.040\textwidth, cfbox=red 1pt 0pt]{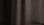} &
        \includegraphics[width=0.115\textwidth, height=0.040\textwidth, cfbox=red 1pt 0pt]{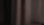} &
        \includegraphics[width=0.115\textwidth, height=0.040\textwidth, cfbox=red 1pt 0pt]{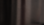} &
        \includegraphics[width=0.115\textwidth, height=0.040\textwidth, cfbox=red 1pt 0pt]{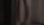} &
        {} &
        \includegraphics[width=0.115\textwidth, height=0.040\textwidth, cfbox=red 1pt 0pt]{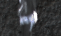} &
        \includegraphics[width=0.115\textwidth, height=0.040\textwidth, cfbox=red 1pt 0pt]{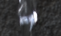} &
        \includegraphics[width=0.115\textwidth, height=0.040\textwidth, cfbox=red 1pt 0pt]{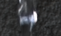} &
        \includegraphics[width=0.115\textwidth, height=0.040\textwidth, cfbox=red 1pt 0pt]{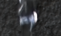} \\

        \small \textit{reflective\_29} &
        \small 37.01/0.9841 &
        \small 37.34/0.9847 &
        \small 36.84/0.9846 &
        {} &
        \small \textit{Water\_Drops} &
        \small 38.01/0.9832 &
        \small 38.81/0.9614 &
        \small 38.73/0.9650 \\

    \end{tabular}
    \caption{Visualization of failed examples of SADenseNet. Error maps are depicted as heat maps in the first row, zoom-in views of selected regions are displayed in the red boxes in the second row, and sample titles and PSNR/SSIM are given in the last row.}
    \label{fig:limitationVis}
\end{figure*}

\subsection{Limitations}
Even though SADenseNet has demonstrated superior performance in the task of LF reconstruction, it doesn't necessarily outperform other methods across all samples. While there are 222 real-world samples tested in the previous evaluation, SADenseNet fails 14 of them, in most of which reflective surfaces are presented. Two of them are visualized in Fig. \ref{fig:limitationVis} with NoisyLFRecon \cite{ZhouNoisyLFRecon_SS2021} and Yeung \textit{et al.} \cite{yeungSAS_ECCV2018} as the representatives of disparity-based and separable-convolution-based methods, respectively.

In \textit{reflective\_29}, it is observed that SADenseNet is outperformed by Yeung \textit{et al.} \cite{yeungSAS_ECCV2018} by 0.5 db PSNR, and in the error maps the reflective surface of the steel pot is reconstructed with distorted artifacts in the red box which doesn't happen in the results of the other two methods. In the other example, \textit{Water\_Drops}, in spite of the higher PSNR and SSIM achieved by Yeung \textit{et al.} \cite{yeungSAS_ECCV2018} and SADenseNet, the flaw can be easily spotted in their error maps that the reflective water drops are not correctly reconstructed, while NoisyLFRecon \cite{ZhouNoisyLFRecon_SS2021} is not confused by these reflections with a visibly clearer error map. Given these results, we presumed that disparity-based methods are more robust to the scenes with reflective surfaces, which remains a problem for our method to study in future research.

\section{Conclusion}
In this paper, We have studied domain asymmetry in LF images and proposed correlation blocks to extract spatio-angular feature representation in an asymmetrical manner. For further enhancing the spatio-angular representation, we have applied spatial and angular dense skip connections to construct a compact information flow. Moreover, additional image skip connections are utilized to complement the correlation features with raw image information, forming our final Spatio-Angular Dense Network (SADenseNet). Experiments on four real-world datasets and a synthetic dataset have demonstrated its state-of-the-art performance at significantly lower costs. Ablation studies have verified the benefits of the skip connections and the efficacy of the proposed asymmetrical processing in the spatial and angular domains using the correlation blocks. Furthermore, we have found that some lightweight variants of our proposed method, \textit{i.e.} the model without spatial dense skip connections or that employing only two correlation blocks, can still achieve impressive performance with very few resources when there are some constrained conditions. At last, we have performed depth estimation as a typical LF measurement task on the reconstructed LF images, and the result has proved SADenseNet's promising potential of improving LF-related measurement performance.


\ifCLASSOPTIONcaptionsoff
\newpage
\fi

\bibliographystyle{IEEEtran}
\bibliography{bare_jrnl}

\begin{thebibliography}{10}
\providecommand{\url}[1]{#1}
\csname url@samestyle\endcsname
\providecommand{\newblock}{\relax}
\providecommand{\bibinfo}[2]{#2}
\providecommand{\BIBentrySTDinterwordspacing}{\spaceskip=0pt\relax}
\providecommand{\BIBentryALTinterwordstretchfactor}{4}
\providecommand{\BIBentryALTinterwordspacing}{\spaceskip=\fontdimen2\font plus
\BIBentryALTinterwordstretchfactor\fontdimen3\font minus
  \fontdimen4\font\relax}
\providecommand{\BIBforeignlanguage}[2]{{%
\expandafter\ifx\csname l@#1\endcsname\relax
\typeout{** WARNING: IEEEtran.bst: No hyphenation pattern has been}%
\typeout{** loaded for the language `#1'. Using the pattern for}%
\typeout{** the default language instead.}%
\else
\language=\csname l@#1\endcsname
\fi
#2}}
\providecommand{\BIBdecl}{\relax}
\BIBdecl

\bibitem{raghavendraLFFace_TIP2015}
R.~Raghavendra, K.~B. Raja, and C.~Busch, ``Presentation attack detection for
  face recognition using light field camera,'' \emph{IEEE Transactions on Image
  Processing}, vol.~24, no.~3, pp. 1060--1075, 2015.

\bibitem{jiLFHOG_ICIP2016}
Z.~Ji, H.~Zhu, and Q.~Wang, ``{{LFHOG}}: {{A}} discriminative descriptor for
  live face detection from light field image,'' in \emph{2016 {{IEEE}}
  International Conference on Image Processing ({{ICIP}})}.\hskip 1em plus
  0.5em minus 0.4em\relax {IEEE}, 2016, pp. 1474--1478.

\bibitem{sellahewaFace_TIM2010}
H.~Sellahewa and S.~A. Jassim, ``Image-quality-based adaptive face
  recognition,'' \emph{IEEE Transactions on Instrumentation and Measurement},
  vol.~59, no.~4, pp. 805--813, 2010.

\bibitem{fangFace_TIM2015}
L.~Fang and S.~Li, ``Face recognition by exploiting local gabor features with
  multitask adaptive sparse representation,'' \emph{IEEE Transactions on
  Instrumentation and Measurement}, vol.~64, no.~10, pp. 2605--2615, 2015.

\bibitem{wangLFRecognition_ECCV2016}
T.-C. Wang, J.-Y. Zhu, E.~Hiroaki, M.~Chandraker, A.~A. Efros, and
  R.~Ramamoorthi, ``A 4d light-field dataset and cnn architectures for material
  recognition,'' in \emph{Computer Vision -- ECCV 2016}.\hskip 1em plus 0.5em
  minus 0.4em\relax Cham: Springer International Publishing, 2016, pp.
  121--138.

\bibitem{luLFRecognition_2019}
Z.~Lu, H.~W.~F. Yeung, Q.~Qu, Y.~Y. Chung, X.~Chen, and Z.~Chen, ``Improved
  image classification with {{4D}} light-field and interleaved convolutional
  neural network,'' \emph{Multimedia Tools and Applications}, vol.~78, no.~20,
  pp. 29\,211--29\,227, 2019-10.

\bibitem{songEDRNet_TIM2020}
G.~Song, K.~Song, and Y.~Yan, ``Edrnet: Encoder–decoder residual network for
  salient object detection of strip steel surface defects,'' \emph{IEEE
  Transactions on Instrumentation and Measurement}, vol.~69, no.~12, pp.
  9709--9719, 2020.

\bibitem{shengLFSaliency_ICASSP2016}
H.~Sheng, S.~Zhang, X.~Liu, and Z.~Xiong, ``Relative location for light field
  saliency detection,'' in \emph{2016 {{IEEE International Conference}} on
  {{Acoustics}}, {{Speech}} and {{Signal Processing}} ({{ICASSP}})}.\hskip 1em
  plus 0.5em minus 0.4em\relax {IEEE}, 2016, pp. 1631--1635.

\bibitem{zhangLFNet_TIP2020}
M.~Zhang, W.~Ji, Y.~Piao, J.~Li, Y.~Zhang, S.~Xu, and H.~Lu, ``Lfnet: {{Light}}
  field fusion network for salient object detection,'' \emph{IEEE Transactions
  on Image Processing}, vol.~29, pp. 6276--6287, 2020.

\bibitem{heberUshapeICCV2017}
S.~Heber, W.~Yu, and T.~Pock, ``Neural {{EPI}}-{{Volume Networks}} for
  {{Shape}} from {{Light Field}},'' in \emph{Proceedings of the {{IEEE
  International Conference}} on {{Computer Vision}}}, vol. 2017-Octob, Oct.
  2017, pp. 2271--2279.

\bibitem{jeonDepthLightField2019}
H.-G. Jeon, J.~Park, G.~Choe, J.~Park, Y.~Bok, Y.-W. Tai, and I.~S. Kweon,
  ``Depth from a {{Light Field Image}} with {{Learning}}-{{Based Matching
  Costs}},'' \emph{IEEE Transactions on Pattern Analysis and Machine
  Intelligence}, vol.~41, no.~2, pp. 297--310, Feb. 2019.

\bibitem{wangOcclusionawareDepthEstimation2015}
T.-C. Wang, A.~A. Efros, and R.~Ramamoorthi, ``Occlusion-aware depth estimation
  using light-field cameras,'' in \emph{Proceedings of the {{IEEE International
  Conference}} on {{Computer Vision}}}, 2015, pp. 3487--3495.

\bibitem{chen2018accurate}
J.~Chen, J.~Hou, Y.~Ni, and L.-P. Chau, ``Accurate light field depth estimation
  with superpixel regularization over partially occluded regions,'' \emph{IEEE
  Transactions on Image Processing}, vol.~27, no.~10, pp. 4889--4900, 2018.

\bibitem{linStereoDepth_TIM2021}
H.-Y. Lin, C.-L. Tsai, and V.~L. Tran, ``Depth measurement based on stereo
  vision with integrated camera rotation,'' \emph{IEEE Transactions on
  Instrumentation and Measurement}, vol.~70, pp. 1--10, 2021.

\bibitem{lilienblum3DStructuredLight_TIM2014}
E.~Lilienblum and A.~Al-Hamadi, ``A structured light approach for 3-d surface
  reconstruction with a stereo line-scan system,'' \emph{IEEE Transactions on
  Instrumentation and Measurement}, vol.~64, no.~5, pp. 1258--1266, 2015.

\bibitem{website:Raytrix}
\BIBentryALTinterwordspacing
``Raytrix \textendash{} {{3D}} light field camera technology.'' [Online].
  Available: \url{https://raytrix.de/}
\BIBentrySTDinterwordspacing

\bibitem{heinze2016automated}
C.~Heinze, S.~Spyropoulos, S.~Hussmann, and C.~Perwa{\ss}, ``Automated robust
  metric calibration algorithm for multifocus plenoptic cameras,'' \emph{IEEE
  Transactions on Instrumentation and Measurement}, vol.~65, no.~5, pp.
  1197--1205, 2016.

\bibitem{website:Lytro}
\BIBentryALTinterwordspacing
``{Lytro}.'' [Online]. Available: \url{https://www.lytro.com/}
\BIBentrySTDinterwordspacing

\bibitem{deflectometry2021}
Z.~{Niu}, X.~{Zhang}, J.~{Ye}, L.~{Yea}, R.~{Zhu}, and X.~{Jiang}, ``Adaptive
  phase correction for phase measuring deflectometry based on light field
  modulation,'' \emph{IEEE Transactions on Instrumentation and Measurement},
  pp. 1--1, 2021.

\bibitem{kalantari_SIGGRAPHASIA2016}
N.~K. Kalantari, T.-C. Wang, and R.~Ramamoorthi, ``Learning-based view
  synthesis for light field cameras,'' \emph{ACM Transactions on Graphics
  (Proceedings of SIGGRAPH Asia 2016)}, vol.~35, no.~6, p. 193, 2016.

\bibitem{yeungSAS_ECCV2018}
H.~Wing Fung~Yeung, J.~Hou, J.~Chen, Y.~Y. Chung, and X.~Chen, ``Fast {{Light
  Field Reconstruction With Deep Coarse}}-{{To}}-{{Fine Modeling}} of
  {{Spatial}}-{{Angular Clues}},'' in \emph{The {{European Conference}} on
  {{Computer Vision}} ({{ECCV}})}, Sep. 2018.

\bibitem{wuEPICNN_CVPR2017}
G.~Wu, M.~Zhao, L.~Wang, Q.~Dai, T.~Chai, and Y.~Liu, ``Light field
  reconstruction using deep convolutional network on {{EPI}},'' in \emph{{{IEEE
  Conference}} on {{Computer Vision}} and {{Pattern Recognition}} ({{CVPR}})},
  vol. 2017, 2017, p.~2.

\bibitem{yeungSAS_TIP2019}
H.~W.~F. Yeung, J.~Hou, X.~Chen, J.~Chen, Z.~Chen, and Y.~Y. Chung, ``Light
  {{Field Spatial Super}}-{{Resolution Using Deep Efficient Spatial}}-{{Angular
  Separable Convolution}},'' \emph{IEEE Transactions on Image Processing},
  vol.~28, no.~5, pp. 2319--2330, 2019.

\bibitem{wuEPICNN_TPAMI2018}
G.~Wu, Y.~Liu, L.~Fang, Q.~Dai, and T.~Chai, ``Light {{Field Reconstruction
  Using Convolutional Network}} on {{EPI}} and {{Extended Applications}},''
  \emph{IEEE Transactions on Pattern Analysis and Machine Intelligence}, 2018.

\bibitem{shinEPINET2018}
C.~Shin, H.-G. Jeon, Y.~Yoon, I.~S. Kweon, and S.~J. Kim, ``Epinet: A
  fully-convolutional neural network using epipolar geometry for depth from
  light field images,'' in \emph{Proceedings of the IEEE Conference on Computer
  Vision and Pattern Recognition}, 2018, pp. 4748--4757.

\bibitem{heberUshapeBMVC2016}
S.~Heber, W.~Yu, and T.~Pock, ``U-shaped {{Networks}} for {{Shape}} from
  {{Light Field}},'' in \emph{Procedings of the {{British Machine Vision
  Conference}} 2016}, vol.~1, 2016, pp. 37.1--37.12.

\bibitem{wangPseudo4DCNN2018}
Y.~Wang, F.~Liu, Z.~Wang, G.~Hou, Z.~Sun, and T.~Tan, ``End-to-end {View}
  {Synthesis} for {Light} {Field} {Imaging} with {Pseudo} 4dcnn,'' in \emph{The
  {European} {Conference} on {Computer} {Vision} ({ECCV})}, Sep. 2018.

\bibitem{simonyanVGG_2015}
K.~Simonyan and A.~Zisserman, ``Very {{Deep Convolutional Networks}} for
  {{Large}}-{{Scale Image Recognition}},'' \emph{International Conference on
  Learning Representations (ICRL)}, pp. 1--14, 2015.

\bibitem{kim_accurate_2016}
J.~Kim, J.~K. Lee, and K.~M. Lee, ``Accurate {Image} {Super}-{Resolution}
  {Using} {Very} {Deep} {Convolutional} {Networks},'' in \emph{{IEEE}
  {Computer} {Society} {Conference} on {Computer} {Vision} and {Pattern}
  {Recognition} ({CVPR})}, 2016, pp. 1646--1654.

\bibitem{huangDenseNet_CVPR2017}
G.~Huang, Z.~Liu, L.~van~der Maaten, and K.~Q. Weinberger, ``Densely
  {{Connected Convolutional Networks}},'' in \emph{Proceedings of the {{IEEE
  Conference}} on {{Computer Vision}} and {{Pattern Recognition}}}, 2017, pp.
  4700--4708.

\bibitem{taiMemNet_ICCV2017}
Y.~Tai, J.~Yang, X.~Liu, and C.~Xu, ``{{MemNet}}: {{A Persistent Memory
  Network}} for {{Image Restoration}},'' in \emph{Proceedings of the {{IEEE}}
  International Conference on Computer Vision}, 2017, pp. 4539--4547.

\bibitem{tongSRDenseNet_ICCV2017}
T.~Tong, G.~Li, X.~Liu, and Q.~Gao, ``Image {{Super}}-{{Resolution Using Dense
  Skip Connections}},'' in \emph{2017 {{IEEE International Conference}} on
  {{Computer Vision}} ({{ICCV}})}.\hskip 1em plus 0.5em minus 0.4em\relax
  {IEEE}, 2017, pp. 4809--4817.

\bibitem{harisDBPN_CVPR2018}
M.~Haris, G.~Shakhnarovich, and N.~Ukita, ``Deep back-projection networks for
  super-resolution,'' in \emph{Proceedings of the {IEEE} conference on computer
  vision and pattern recognition}, 2018, pp. 1664--1673.

\bibitem{liSRFBN_CVPR2019}
Z.~Li, J.~Yang, Z.~Liu, X.~Yang, G.~Jeon, and W.~Wu, ``Feedback {{Network}} for
  {{Image Super}}-{{Resolution}},'' in \emph{Proceedings of the {{IEEE
  Conference}} on {{Computer Vision}} and {{Pattern Recognition}}}, 2019, pp.
  3867--3876.

\bibitem{pujadesBayesianViewSynthesis2014}
S.~Pujades, F.~Devernay, and B.~Goldluecke, ``Bayesian view synthesis and
  image-based rendering principles,'' in \emph{Proceedings of the {{IEEE
  Conference}} on {{Computer Vision}} and {{Pattern Recognition}}}, 2014, pp.
  3906--3913.

\bibitem{zhang2015light}
Z.~Zhang, Y.~Liu, and Q.~Dai, ``Light field from micro-baseline image pair,''
  in \emph{Proceedings of the IEEE Conference on Computer Vision and Pattern
  Recognition}, 2015, pp. 3800--3809.

\bibitem{wanner2014variational}
S.~Wanner and B.~Goldluecke, ``Variational light field analysis for disparity
  estimation and super-resolution,'' \emph{IEEE transactions on pattern
  analysis and machine intelligence}, vol.~36, no.~3, pp. 606--619, 2014.

\bibitem{penner_soft_2017}
E.~Penner and L.~Zhang, ``Soft 3d reconstruction for view synthesis,''
  \emph{ACM Transactions on Graphics (TOG)}, vol.~36, no.~6, p. 235, 2017.

\bibitem{zhengCrossNet_ECCV2018}
H.~Zheng, M.~Ji, H.~Wang, Y.~Liu, and L.~Fang, ``{{CrossNet}}: {{An
  End}}-to-end {{Reference}}-based {{Super Resolution Network}} using
  {{Cross}}-scale {{Warping}},'' in \emph{Proceedings of the {{European
  Conference}} on {{Computer Vision}} ({{ECCV}})}, 2018, pp. 88--104.

\bibitem{pengLargeKernelGCN_CVPR2017}
C.~Peng, X.~Zhang, G.~Yu, G.~Luo, and J.~Sun, ``Large kernel matters -
  {{Improve}} semantic segmentation by global convolutional network,'' in
  \emph{Proceedings - 30th {{IEEE Conference}} on {{Computer Vision}} and
  {{Pattern Recognition}}, {{CVPR}} 2017}, vol. 2017-Janua, 2017, pp.
  1743--1751.

\bibitem{caballeroVESPCN_CVPR2017}
J.~Caballero, C.~Ledig, A.~Aitken, A.~Acosta, J.~Totz, Z.~Wang, and W.~Shi,
  ``Real-time video super-resolution with spatio-temporal networks and motion
  compensation,'' in \emph{Proceedings of the {{IEEE Conference}} on {{Computer
  Vision}} and {{Pattern Recognition}}}, 2017, pp. 4778--4787.

\bibitem{ZhouNoisyLFRecon_SS2021}
W.~Zhou, J.~Shi, Y.~Hong, L.~Lin, and E.~E. Kuruoglu, ``Robust dense light
  field reconstruction from sparse noisy sampling,'' \emph{Signal Processing},
  vol. 186, p. 108121, 2021.

\bibitem{szegedyGoogLeNet_CVPR2015}
C.~Szegedy, W.~Liu, Y.~Jia, P.~Sermanet, S.~Reed, D.~Anguelov, D.~Erhan,
  V.~Vanhoucke, and A.~Rabinovich, ``Going deeper with convolutions,'' in
  \emph{Proceedings of the {{IEEE}} Conference on Computer Vision and Pattern
  Recognition}, 2015, pp. 1--9.

\bibitem{mengHDDRNet_TPAMI2019}
N.~Meng, H.~K.-H. So, X.~Sun, and E.~Lam, ``High-dimensional dense residual
  convolutional neural network for light field reconstruction,'' \emph{{IEEE}
  Transactions on Pattern Analysis and Machine Intelligence}, 2019.

\bibitem{ChandramouliGenerativeLF_TPAMI2020}
P.~Chandramouli, K.~V. Gandikota, A.~Gorlitz, A.~Kolb, and M.~Moeller, ``A
  {{Generative Model}} for {{Generic Light Field Reconstruction}},'' \emph{IEEE
  Transactions on Pattern Analysis and Machine Intelligence}, 2020.

\bibitem{GuoLFCA_ECCV2020}
``Deep spatial-angular regularization for compressive light field
  reconstruction over coded apertures,'' in \emph{Computer Vision -- ECCV
  2020}.\hskip 1em plus 0.5em minus 0.4em\relax Cham: Springer International
  Publishing, 2020, pp. 278--294.

\bibitem{InagakiCodedAperture_ECCV2018}
Y.~Inagaki, Y.~Kobayashi, K.~Takahashi, T.~Fujii, and H.~Nagahara, ``Learning
  to capture light fields through a coded aperture camera,'' in
  \emph{Proceedings of the {{European Conference}} on {{Computer Vision}}
  ({{ECCV}})}, 2018, pp. 418--434.

\bibitem{heResNet_CVPR2016}
K.~He, X.~Zhang, S.~Ren, and J.~Sun, ``Deep {Residual} {Learning} for {Image}
  {Recognition},'' in \emph{2016 {IEEE} {Conference} on {Computer} {Vision} and
  {Pattern} {Recognition} ({CVPR})}, 2016, pp. 770--778.

\bibitem{tai_image_2017}
Y.~Tai, J.~Yang, and X.~Liu, ``Image super-resolution via deep recursive
  residual network,'' in \emph{Proceedings of the {IEEE} conference on computer
  vision and pattern recognition}, 2017, pp. 3147--3155.

\bibitem{zhangRCAN_ECCV2018}
Y.~Zhang, K.~Li, K.~Li, L.~Wang, B.~Zhong, and Y.~Fu, ``Image super-resolution
  using very deep residual channel attention networks,'' in \emph{European
  {Conference} on {Computer} {Vision}}, 2018, pp. 286--301.

\bibitem{songCREST_ICCV2017}
Y.~Song, C.~Ma, L.~Gong, J.~Zhang, R.~W.~H. Lau, and M.-H. Yang, ``{{CREST}}:
  {{Convolutional Residual Learning}} for {{Visual Tracking}},'' in
  \emph{Proceedings of the {{IEEE International Conference}} on {{Computer
  Vision}}}, 2017, pp. 2555--2564.

\bibitem{wang_learning_2018}
Q.~Wang, Z.~Teng, J.~Xing, J.~Gao, W.~Hu, and S.~Maybank, ``Learning
  {Attentions}: {Residual} {Attentional} {Siamese} {Network} for {High}
  {Performance} {Online} {Visual} {Tracking},'' in \emph{{IEEE} {Computer}
  {Society} {Conference} on {Computer} {Vision} and {Pattern} {Recognition}
  ({CVPR})}, 2018.

\bibitem{qiuP3D_ICCV2017}
Z.~Qiu, T.~Yao, and T.~Mei, ``Learning {Spatio}-{Temporal} {Representation}
  {With} {Pseudo}-3d {Residual} {Networks},'' in \emph{Proceedings of the
  {IEEE} {International} {Conference} on {Computer} {Vision}}, 2017, pp.
  5533--5541.

\bibitem{liP3DSR_CVPR2019}
S.~Li, F.~He, B.~Du, L.~Zhang, Y.~Xu, and D.~Tao, ``Fast {Spatio}-{Temporal}
  {Residual} {Network} for {Video} {Super}-{Resolution},'' in \emph{Proceedings
  of the {IEEE} {Conference} on {Computer} {Vision} and {Pattern}
  {Recognition}}, 2019, pp. 10\,522--10\,531.

\bibitem{tranR2D_CVPR2018}
D.~Tran, H.~Wang, L.~Torresani, J.~Ray, Y.~LeCun, and M.~Paluri, ``A closer
  look at spatiotemporal convolutions for action recognition,'' in
  \emph{Proceedings of the {IEEE} conference on {Computer} {Vision} and
  {Pattern} {Recognition}}, 2018, pp. 6450--6459.

\bibitem{ilgFlowNet2_CVPR2017}
E.~Ilg, N.~Mayer, T.~Saikia, M.~Keuper, A.~Dosovitskiy, and T.~Brox,
  ``{{FlowNet}} 2.0: {{Evolution}} of {{Optical Flow Estimation}} with {{Deep
  Networks}},'' in \emph{2017 {{IEEE Conference}} on {{Computer Vision}} and
  {{Pattern Recognition}} ({{CVPR}})}, Jul. 2017, pp. 1647--1655.

\bibitem{wannerHCI2013}
S.~Wanner, S.~Meister, and B.~Goldluecke, ``Datasets and benchmarks for densely
  sampled 4d light fields.'' in \emph{{{VMV}}}.\hskip 1em plus 0.5em minus
  0.4em\relax {Citeseer}, 2013, pp. 225--226.

\bibitem{keras}
F.~Chollet \emph{et~al.}, ``Keras,'' \url{https://keras.io}, 2015.

\bibitem{tensorflow}
M.~Abadi, P.~Barham, J.~Chen, Z.~Chen, A.~Davis, J.~Dean, M.~Devin,
  S.~Ghemawat, G.~Irving, M.~Isard \emph{et~al.}, ``Tensorflow: A system for
  large-scale machine learning,'' in \emph{12th $USENIX$ Symposium on Operating
  Systems Design and Implementation $OSDI$ 16)}, 2016, pp. 265--283.

\bibitem{kingmaAdam_2014}
D.~P. Kingma and J.~Ba, ``Adam: {{A}} method for stochastic optimization,''
  \emph{arXiv preprint arXiv:1412.6980}, 2014.

\bibitem{rerabekEPFL2016}
M.~Rerábek and T.~Ebrahimi, ``New {{Light Field Image Dataset}},'' \emph{8th
  International Conference on Quality of Multimedia Experience (QoMEX)}, pp.
  1--2, 2016.

\bibitem{StanfordLytro}
\BIBentryALTinterwordspacing
``Stanford {{Lytro Light Field Archive}}.'' [Online]. Available:
  \url{http://lightfields.stanford.edu/LF2016.html}
\BIBentrySTDinterwordspacing

\end{thebibliography}

\begin{IEEEbiography}[{\includegraphics[width=1in,keepaspectratio]{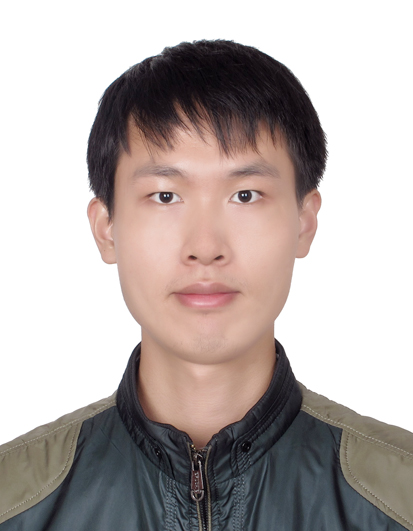}}]{Zexi Hu} received his B.S. degree in Software Engineering from South China Agricultural University in 2014; and his M.Phil. degree in Engineering and IT from the University of Sydney in 2019. He is an invited expert of the Multimedia Systems Lab, Beijing Technology and Business University (BTBU). His research interests include computer vision and deep learning.
\end{IEEEbiography}

\vspace{-30pt}

\begin{IEEEbiography}[{\includegraphics[width=1in,keepaspectratio]{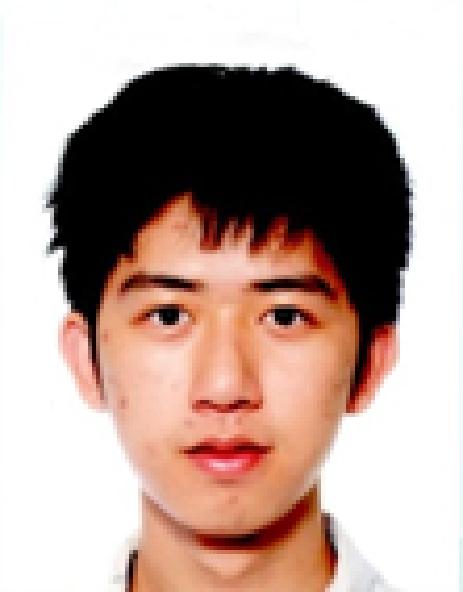}}] {Henry Wing Fung Yeung} received his M.Phil. and Ph.D. degrees in Engineering and IT from University of Sydney. He has authored or co-authored multiple international journal and conference papers, including the European Conference on Computer Vision and IEEE Trans. on Image Processing. His research interests include light field image processing and machine learning.
\end{IEEEbiography}

\vspace{-30pt}

\begin{IEEEbiography}[{\includegraphics[width=1in,keepaspectratio]{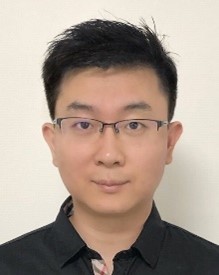}}] {Xiaoming Chen} received the B.Sc. degree from the Royal Melbourne Institute of Technology, Australia, in 2004, and the Ph.D. degree from the University of Sydney, Australia, in 2009. From 2010 to 2014, he was with the National University of Singapore, CSIRO Australia, and IBM. 

From 2014 to 2019, he was a Researcher with the Institute of Advanced Technology, University of Science and Technology of China. In 2014, he was invited into the “100-Talent Program” by the Government of Hefei, Anhui Province, China. He is currently a Professor with the School of Computer Science and Engineering, Beijing Technology and Business University (BTBU), China. Meanwhile, he has been invited as a Guest Researcher with Beijing Research Institute, University of Science and Technology of China since 2019, and a Guest Researcher at the School of Computer Science, the University of Sydney. His research interests include immersive media computing, virtual reality, and related business information systems and applications. His work has been published in journals and conferences including IEEE Trans. on Image Processing, IEEE Trans. on Circuits and Systems for Video Technology, IEEE Trans. on Broadcasting, IEEE Trans. on Instrumentation and Measurement, IEEE Multimedia, ACM Multimedia, IEEE Virtual Reality, European Conference on Computer Vision, etc.
\end{IEEEbiography}

\vspace{-30pt}

\begin{IEEEbiography}[{\includegraphics[width=1in,keepaspectratio]{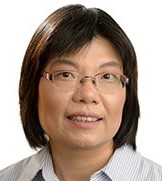}}] {Yuk Ying Chung} (M’05, M’17) received her B.S. degree in computing and information systems from University of London, UK, in 1995; and her Ph.D. degree in computer engineering from the Queensland University of Technology, Australia in 2000. 

From 1999 to 2001, she was a lecturer at La Trobe University in Melbourne, Australia. She is a Senior Lecturer and has been with the School of Computer Science, University of Sydney, Australia since 2001. Her research interests include image and video processing, virtual reality, deep neural network, machine learning and data mining. Her work has been published in conferences and journals including ECCV, NeurIPS, IEEE Trans. on Image Processing, and IEEE Trans. on Circuits and Systems for Video Technology, IEEE Trans. on Cybernetics, etc.

\end{IEEEbiography}

\vspace{-30pt}

\begin{IEEEbiography}[{\includegraphics[width=1in,keepaspectratio]{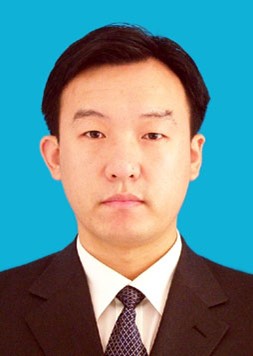}}] {Haisheng Li} received the Ph.D. degree in computer graphics from Beihang University, Beijing, China, in 2002. He is currently a Professor and Dean with the School of Computer Science and Engineering, Beijing Technology and Business University (BTBU), China. His current research interests include computer graphics, scientific visualization, 3-D model retrieval.
\end{IEEEbiography}

\end{document}